\documentclass[sigconf, nonacm]{acmart}

\usepackage{graphicx}
\usepackage{hyperref}
\usepackage{balance}  % for  \balance command ON LAST PAGE  (only there!)

\usepackage{color}
\usepackage{multirow}
\usepackage{subfigure}
\usepackage[export]{adjustbox}

\newcommand{\stitle}[1]{\vspace{0.8ex}\noindent\textup{\textbf{#1}}}

\usepackage{algorithmicx}
\usepackage{algorithm}
\usepackage{algpseudocode}

\newtheorem{definition}{Definition}

\newcommand{\hide}[1]{}

\def\bitcoinA{%
  \leavevmode
  \vtop{\offinterlineskip %\bfseries
    \setbox0=\hbox{B}%
    \setbox2=\hbox to\wd0{\hfil\hskip-.03em
    \vrule height .3ex width .15ex\hskip .08em
    \vrule height .3ex width .15ex\hfil}
  \vbox{\copy2\box0}\box2}}

\def\bitcoinB{\leavevmode
  {\setbox0=\hbox{\textsf{B}}%
    \dimen0\ht0 \advance\dimen0 0.2ex
    \ooalign{\hfil \box0\hfil\cr
      \hfil\vrule height \dimen0 depth.2ex\hfil\cr
    }%
  }%
}

\pdfoutput=1

\makeatletter
	\def\@copyrightspace{\relax}
	\makeatother

\begin{document}
\title{Provenance in Temporal Interaction Networks}

%%
%% The "author" command and its associated commands are used to define the authors and their affiliations.
% 1st. author
  \author{Chrysanthi Kosyfaki, Nikos Mamoulis}
  \affiliation{
       \institution{Department of Computer Science and Engineering}
       \institution{University of Ioannina}
     }
    \email{{xkosifaki,nikos}@cse.uoi.gr}
     
  %\author{Nikos Mamoulis}
  %\affiliation{
      % \institution{Department of Computer Science and Engineering}
      % \institution{University of Ioannina}
     %}
    %\email{nikos@cse.uoi.gr}


\begin{abstract}
In temporal interaction networks, vertices correspond to entities,
which exchange data quantities
(e.g., money, bytes, messages) over time.
Tracking the origin of data that have reached a given vertex at any 
time can help data analysts to understand the reasons
behind the accumulated quantity at the vertex or
behind the interactions between entities.
%\nikos{might need to rephrase this}
%are often defined as networks which consider not
%only the information (flow) transfers among the vertices but also the
%exact time of the transaction. It is vital, for these networks, to
%know the origin of this quantity to answer questions related to the
%importance of the nodes and hence to characterize them.
In this paper, we study data provenance
in a temporal interaction network.
We investigate alternative propagation models that may apply to
different application scenarios. 
For each such model, we propose
annotation mechanisms
that track the origin of propagated data in the network
and 
%trace 
the routes of data quantities.
Besides analyzing the space and time complexity of these
mechanisms, we propose techniques that reduce their cost in practice,
by either (i) limiting provenance tracking to a subset of
vertices or groups of vertices, or (ii) tracking provenance only for
quantities that were generated in the near past or limiting the
provenance data in each vertex by a budget constraint. 
Our experimental evaluation on five real datasets shows that
quantity propagation models based on generation time or receipt order
scale well on large graphs; on the other hand, a model that propagates
quantities proportionally has high space and time requirements and
can benefit from the aforementioned cost reduction techniques.

\end{abstract}

\maketitle

%\pagestyle{\vldbpagestyle}
%\begingroup\small\noindent\raggedright\textbf{PVLDB Reference Format:}\\
%\vldbauthors. \vldbtitle. PVLDB, \vldbvolume(\vldbissue): \vldbpages, \vldbyear.\\
%\href{https://doi.org/\vldbdoi}{doi:\vldbdoi}
%\endgroup
%\begingroup
%\renewcommand\thefootnote{}\footnote{\noindent
%This work is licensed under the Creative Commons BY-NC-ND 4.0 International License. Visit \url{https://creativecommons.org/licenses/by-nc-nd/4.0/} to view a copy of this license. For any use beyond those covered by this license, obtain permission by emailing \href{mailto:info@vldb.org}{info@vldb.org}. Copyright is held by the owner/author(s). Publication rights licensed to the VLDB Endowment. \\
%\raggedright Proceedings of the VLDB Endowment, Vol. \vldbvolume, No. \vldbissue\ %
%ISSN 2150-8097. \\
%\href{https://doi.org/\vldbdoi}{doi:\vldbdoi} \\
%}\addtocounter{footnote}{-1}\endgroup

%\ifdefempty{\vldbavailabilityurl}{}{
%\vspace{.3cm}
%\begingroup\small\noindent\raggedright\textbf{PVLDB Artifact Availability:}\\
%The source code, data, and/or other artifacts have been made available at \url{\vldbavailabilityurl}.
%\endgroup

\section{Introduction}\label{sec:intro}
Many real world applications can be represented as {\em temporal
  interaction networks} (TINs) \cite{holme2012temporal},
where vertices correspond to entities or hubs that exchange data over time.
Examples of such graphs are financial exchange networks, road networks, social networks, communication networks, etc.
%, food webs, etc.
Each interaction $r$ in a TIN is characterized by 
a source vertex $r.s$, a destination vertex $r.d$, a timestamp $r.t$ and a quantity $r.q$ (e.g., money, passengers, messages, kbytes, etc.) transferred at time $r.t$.
%Hence, each edge of the network connecting vertex $v$ to vertex $u$
%can be seen as a {\em time-series of interactions} from $v$ to $u$.
% An interaction network carries flow of data (money, messages, kbytes, the number of vehicles on a road) which transferred among the nodes on a specific time period.

%Figure \ref{fig:intro1} shows an example of a cryptocurrency 
%network, where vertices correspond to bitcoin addresses and the
%interactions to transactions between them.
%Each transaction $\langle r.s, r.d, r.t, r.q\rangle$ is viewed as a
%$(r.t, r.q)$ pair on edge $r.s \to r.d$.

%The edges are annotated by the $(r.t,r.q)$ pairs of the corresponding interactions.
%\nikos{change order of numbers in the figure. In each parenthesis,
%  time should be before quantity. For example, (5,2) should become
%  (2,5), tuple (3,4) should become (4,3), etc.} 
% exchange messages. The users on the network represent the nodes and the edges are the interactions happen between the nodes. Each edge is annonated by the time $t$ of the interaction and the quantity of the flow $q$ that is transferred from a user $u$ to a user $v$. In this case, this quantity represents the number of the messages exchanged along a timeline.
% When a vertex $v$ transfers some quantity

\begin{figure}[h]
   \centering
   \includegraphics[width=0.75\columnwidth]{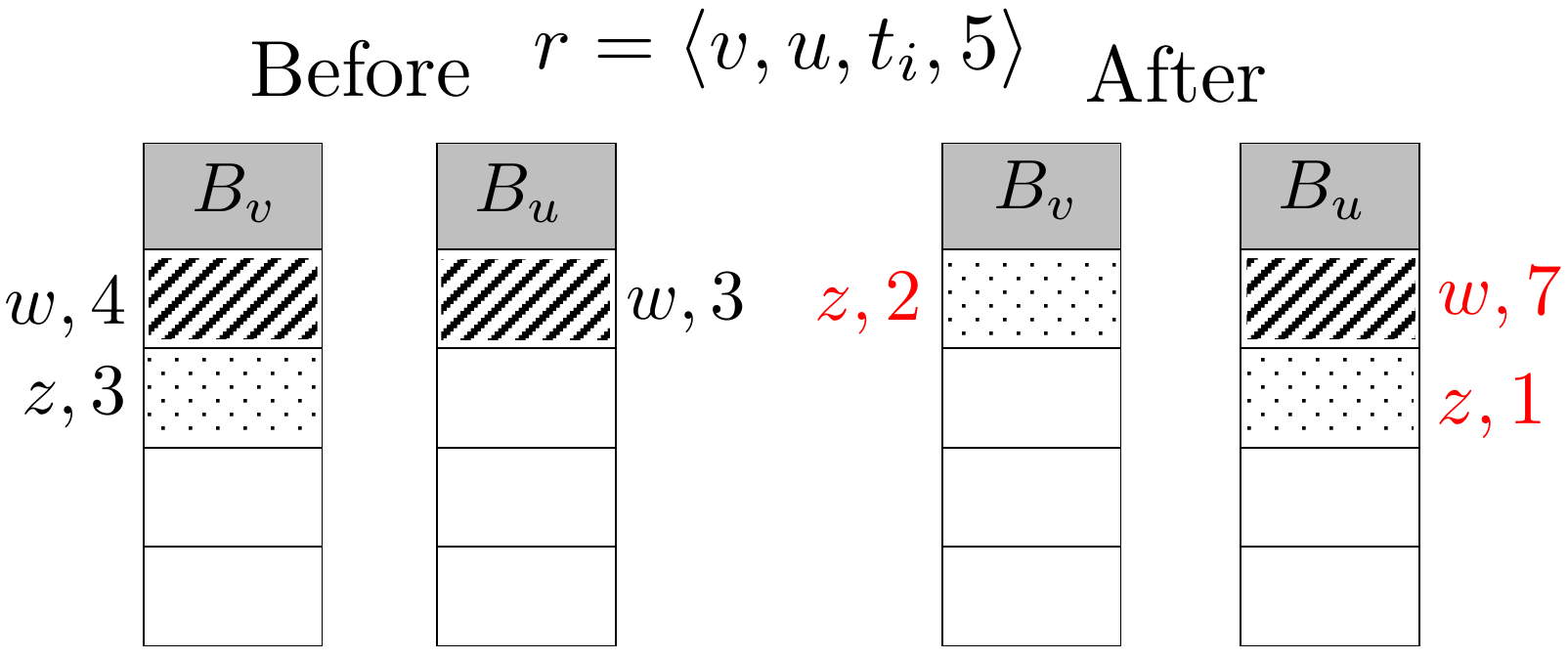}
   \caption{Example of quantity transfer (FIFO policy)}
   \label{fig:intro1}
 \end{figure}
 
\stitle{Objective and Methodology}
%Interaction networks have been studied extensively in the literature.
In this paper, we study a provenance problem in TINs. Our goal is to
track the origin (source) of the quantities that are accumulated at
the vertices over time. To do so, we first need to specify
%define concrete
%models about
how data are transferred from one vertex to another and
when new quantities are generated.
In general, we assume that each
vertex $v$ has a {\em buffer} $B_v$ (e.g., bitcoin wallet) \cite{DBLP:conf/p2p/DeckerW13},
wherein it accumulates all incoming quantities to $v$. Naturally, the buffer $B_v$ changes over time.
Specifically, each interaction $r$ from a vertex $r.v$ to a vertex $r.u$ {\em transfers} $r.q$ units from $B_{r.v}$ to $B_{r.u}$ at time $r.t$.
If $B_{r.v}$ has less than $q$ units by time $r.t$, then the difference must be {\em generated} at $r.v$ before being transferred to $r.u$. 
In a financial exchange network, generation of new quantities could mean that new assets are brought into the network from external sources (e.g., a user buys bitcoins by paying USD). In a road network, new quantities are cars entering the network from a given location.

%\stitle{Overview}
%Our goal is to process the TIN, such that at the end of the timeline we know for each vertex $v$ (i)  the total quantity that is accumulated to $v$ and (ii) where from this quantity originates.
We propose solutions that  {\em proactively} create and propagate lightweight
provenance information in the TIN for the generated quantities, as
they are transferred through the network. This way, we can obtain the origins of the quantities at vertices {\em at any time}.
We define and study 
alternative information propagation models that may apply to
different application scenarios.
Specifically,  consider an interaction $r$ from a vertex $r.v$ to a
vertex $r.u$, which transfers $r.q$ units and assume that $r.q$ is
smaller than the current number of units in buffer $B_{r.v}$.
In this case, $r.q$ units should be {\em selected} from the buffer to
be transferred, based on a policy. The selection policy could
prioritize quantities based on the time they were first generated at their origins, or
on the order they were added to $B_{r.v}$, or could select quantities
proportionally based on their origins.
For instance, Figure \ref{fig:intro1} shows the buffers $B_v$ and $B_u$ of two vertices $v$ and $u$ before and after an interaction $\langle v,u,t_i,5 \rangle$.
The quantities in the buffers are organized as a FIFO queue based on their origins (e.g., $B_v$ contains 4 and 3 units originating from vertex $w$ and vertex $z$, respectively). Based on the FIFO policy, all 4 units from $w$ are selected to be transferred plus 1 unit from $z$. 
For each of the selection policies that we consider, we propose provenance update mechanisms and study
efficient and space-economic algorithms for annotation
propagation.

%Generated quantities can be {\em annotated}, so that their provenance can be traced back, as they are relayed from vertex to vertex.
%at the beginning of the timeline,  
%of use the interaction networks to solve a twofold problem: the problem of flow propagation and its provenance in graphs.

%We identify and study different {\em selection models} for the
%propagation of quantities, based on which
%the origins of the quantities that reach a given vertex may vary.

\stitle{Previous Work}
To our knowledge, there is no previous work that studies data
provenance in TINs. Within our framework, we define and use data
transfer models for TINs, which are based on {\em data relay},
i.e., data units that move in the network are not cloned or deleted.
On the other hand,
most previous works  define and study
information diffusion models \cite{DBLP:conf/kdd/RichardsonD02, DBLP:conf/edbt/0002C17}, where data
items (e.g., news, rumors, etc.) are spread in the network and the
main objective is to identify the vertices of maximal influence in the
network \cite{DBLP:conf/kdd/KempeKT03}, targeting applications such as viral marketing.
Hence, previous efforts on provenance tracing for social networks
\cite{barbier2013provenance,DBLP:conf/www/TaxidouNVFMW15} are based on
different information propagation models compared to our work.

Data provenance is a core concept in
database query evaluation \cite{DBLP:conf/icdt/BunemanKT01} and 
workflow graphs
\cite{DBLP:conf/sigmod/ChapmanJR08, DBLP:journals/pvldb/PsallidasW18}.
The main motivation is 
tracing the 
raw data which contribute to a query output. 
%The graph in this case is the query evaluation plan. %query results are anal
% due to the fact that it is considered as necessary for the database for example to know %the origin of something that happens and there has been a lot of reseach in this field
Data provenance finds use in most types of networks (e.g., threat identification in communication networks \cite{HanP0MS20}).
%Millions of  kbytes are transferred among the vertices, but the system may fall apart because of a specific (malicious) transaction. For solving a problem and finding the culpable of it, it is important to find out from where the problem is started.
%Data provenance is defined as the information describing how data has moved through a network of databases.
Data provenance can be categorized into: {\em where}, {\em how} and
{\em why} provenance \cite{DBLP:journals/ftdb/CheneyCT09}.
Where-provenance identifies the raw data which contribute to some output, 
why-provenance identifies the sources (e.g., tuples) that influenced
the output  \cite{DBLP:journals/pvldb/LeeLG20,
  DBLP:journals/pvldb/LeeLG18, DBLP:journals/vldb/LeeLG19},
and how-provenance explains how the input sources contribute to the output.
%Query execution can be thought of as simply copying data elements from some source to some target database and where-provenance identifies these source elements. Why-provenance provides justification for the data elements appearing in the output and how-provenance describes that some parts of the input influenced certain parts of the output. While there is sample previous work on data provenance, to our knowledge, no previous studies in graphs take into consideration the flow on edges and try to find the origin of the transferred quantity to a given node.
Our work focuses on solving both the where- and why-provenance
problem in TINs, i.e., 
find the vertices that contribute to each vertex over time.
%given the buffer $B_v$ of any vertex $v$ at any time $t$, we are interested in finding out which 
%vertices contributed to $B_v$ and what is their contribution by time $t$.
We also extend our solution to support how-provenance, i.e., capture the paths that have been followed by quantities. 
%from their origins until they reached $v$.
Our solutions can shed light to the reasons behind the accumulation of a quantity at a given vertex. 
Key differences to
previous work on data provenance are: (i) in our problem,
any vertex of the graph can be the origin of a quantity
and any vertex can also be the destination of a propagated quantity;
and (ii) we support the maintenance of provenance information in
real-time,
as new interactions take place in a streaming fashion.

\begin{figure}[h]
   \centering
   \includegraphics[width=0.70\columnwidth]{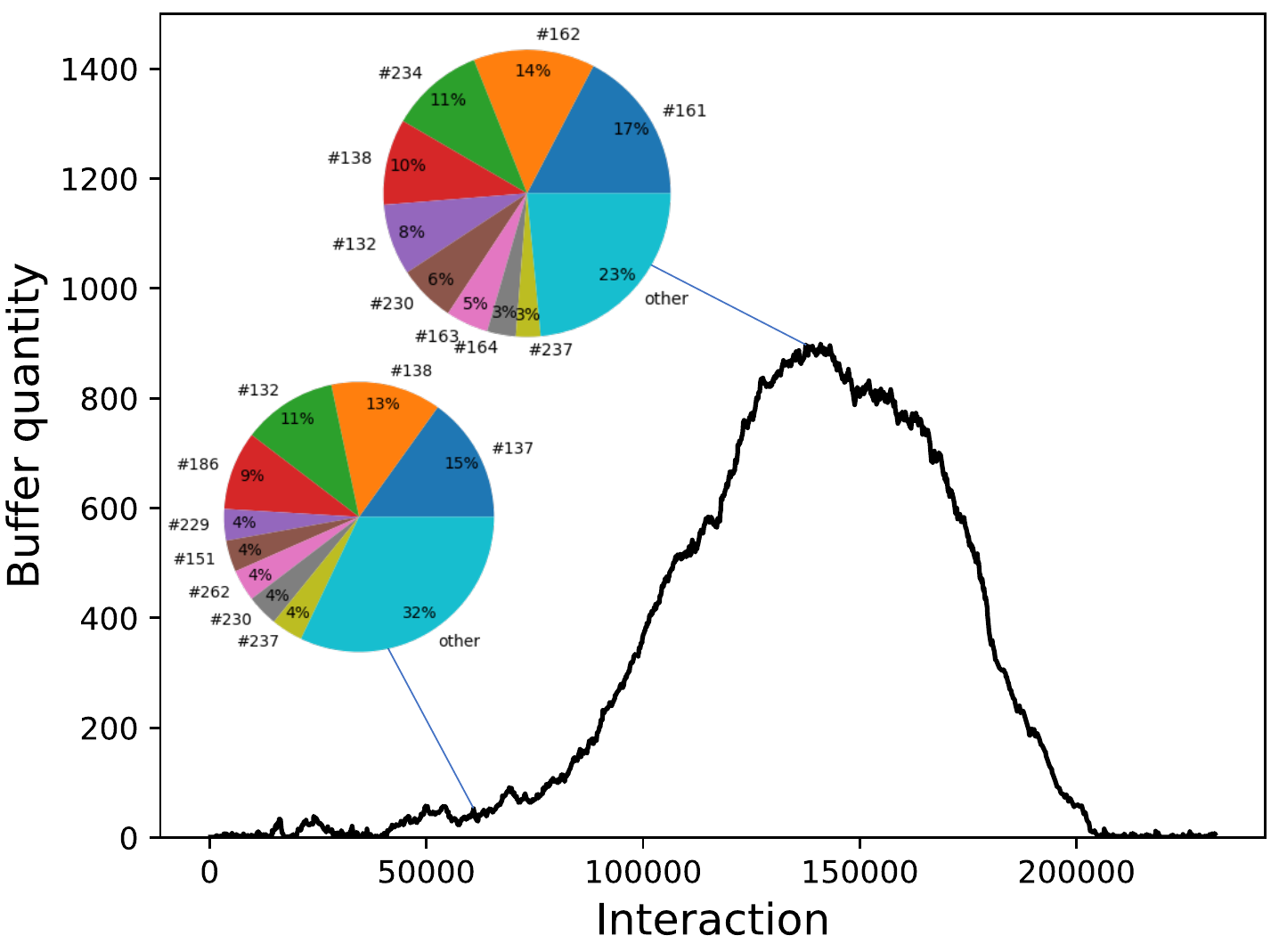}
   \caption{Buffered quantities at vertex \#79 (East Village) after each interaction in Taxis Network}
   \label{fig:intro2}
 \end{figure}

\stitle{Applications}
Solving our provenance problem in TINs finds
application in various domains.
% \cite{DBLP:conf/edbt/ZufleREF18,DBLP:conf/edbt/KosyfakiMPT19,DBLP:conf/icde/KosyfakiMPT21,DBLP:journals/corr/abs-1211-5009}.
In financial networks, tracking the origin of financial units that move 
between accounts can help
in analyzing their relationships. For example, we can identify the accounts that (indirectly) contributed the most in
financing a suspicious account.
%transactions which involve significant transfer of money
%that appears more frequently than expected and may suggest criminal
%behavior).
We can also characterize accounts based on whether they receive
funds from numerous or few sources, or identify groups of users
that finance other groups of users.
%their previous
%activity and their contribution in the final result.
Another example of a TIN is a communication network, where messages
are transferred between vertices and there is a need to trace the origins of malicious
messages that reach a vertex.
Tracing the origin of such messages can be hard due to IP spoofing \cite{MorrisTR} and there is a need for specialized techniques
\cite{DBLP:conf/sigcomm/SavageWKA00}. 
Similarly, in
transportation networks (e.g., flight networks or road networks)
studying the provenance of problems (e.g., traffic, delays, etc.)
may help in finding and alleviating the reasons behind them. 
As a provenance data analysis example, consider one of the TINs used in our experiments, which captures the transfers of passengers by taxi between NYC districts on 2019.01.01. Figure \ref{fig:intro2} shows the number of passengers that are accumulated in East Village from other districts. 
%Observe that this volume gradually increases and starts to drop in the afternoon. 
After each transfer, we can analyze the provenance distribution of passengers
%, i.e., the districts wherefrom they started their trips 
(shown as pie charts).
This can be used to analyze the demographics of visitors over time (e.g., for location-aware marketing).

\stitle{Contributions and outline}
Our main contribution is the formulation of a provenance tracking
problem in temporal interaction networks with important applications.
We define different selection policies for the
propagation of quantities and the corresponding annotation generation
and propagation algorithms.
%Additional contributions of this paper
%include the following:
%\begin{itemize}
%\item
We analyze the space and time complexity of the provenance
  mechanism that we propose for each selection policy and find
  that the {\em proportional propagation policy} is infeasible for large
  graphs because its space complexity is quadratic to the number of
  vertices $|V|$ and each interaction bears a $O(|V|)$ computational
  cost.
  % \item

  We propose restricted, but practical versions of provenance
  tracking under the proportional propagation policy. Our
  {\em selective provenance tracking}
  approach maintains provenance data only from a designated subset of $k$
  vertices, which are of interest to the analyst, 
%  (e.g., the
%  top-$k$ contributing vertices), 
reducing the space complexity to
  $O(k\cdot |V|)$ and the time complexity to $O(k)$ per interaction. 
  The {\em grouped provenance tracking}
  approach tracks provenance from groups of vertices instead of
  individual vertices (e.g., categories of financial entities or
  accounts). Again, the space and time complexity is reduced to
  $O(k\cdot |V|)$ and $O(k)$ per interaction, respectively, if $k$ is
  the number of groups.
  % \item
  We also propose two techniques that limit the scope of provenance
  tracking from all (individual) vertices.
  %, under the proportional
  %propagation policy, given its high complexity.
  The first approach limits provenance tracking up to a
  certain time in the past from the current interaction (i.e., a
  time-window approach). The second approach allocates a
  provenance budget to each vertex.
  Both techniques save
  resources, while providing some guarantees with respect to either
  time or importance of the tracked provenance information.
  % \item

  We extend our propagation algorithms for provenance annotations
  to capture not only the origins of the generated data, but the
  routes (i.e., the paths) that they travelled along in the graph
  until they reached their destinations.
%\item

  We experimentally evaluate the runtime and memory requirements of
  our methods
  on five real TINs with different characteristics. Our results
  show the scalability and limitations of the different  selection
  policies and the corresponding propagation algorithms for provenance data.
%\end{itemize}

The rest of the paper is organized as follows. Section
\ref{sec:rel} reviews related work on data provenance and TINs. In Section
\ref{sec:def}, we formally define the provenance problem in TINs.
Section \ref{sec:models} presents the different information propagation policies
and the corresponding provenance tracking algorithms.
In Section \ref{sec:algorithms}, we discuss
scaleable techniques for provenance
tracking under the proportional propagation policy. 
%Section
%\ref{sec:streaming} proposes techniques for limiting the scope of
%proportional provenance tracking. 
In Section \ref{sec:howprovenance},
we show how to track the paths of the propagated quantities in the
TINs from their origin. Section \ref{sec:exps} presents our
experimental evaluation. Finally, Section \ref{sec:conclusion}
concludes the paper with a discussion about future work.

\section{Related Work}\label{sec:rel}

There has been a lot of research in data provenance over the years \cite{DBLP:journals/pvldb/RupprechtDAGB20,barbier2013provenance,DBLP:journals/pvldb/ChothiaLMR16,DBLP:journals/bmcbi/PaulaHGLW13,DBLP:journals/pvldb/LeeLG20,DBLP:conf/sigmod/Psallidas018,
 DBLP:conf/ipaw/ParulianML21,DBLP:conf/kdd/NamakiFPKAWZW20}. However, we are the first to study the problem of tracking the origin of quantities that flow in temporal interaction networks. In this section, we summarize representative works in temporal interaction networks and provenance tracking.

%Data Provenance is a well studied concept in graph theory and system development \cite{rupprecht2020improving},\cite{barbier2013provenance},\cite{chothia2016explaining}.
%Below we present the most representative works in the field of data provenance.

\subsection{Temporal Interaction Networks}
Temporal interaction networks (TINs) capture the exchange of quantities between entities over time and
they have been studied extensively in the literature
\cite{DBLP:conf/ciac/AkridaCGKS17,li2017fundamental,masuda2013predicting}.
%Analyzing TINs can reveal important information and can be
%used in different domains. 
For instance, Kosyfaki et al. \cite{DBLP:conf/edbt/KosyfakiMPT19} studied the problem of finding recurrent flow path patterns in a TIN, during time windows of specific length.
Pattern detection in TINs with an application in social network analysis was also studied in  \cite{DBLP:conf/edbt/ZufleREF18,DBLP:conf/kdd/BelthZK20}.
A related problem is how to measure the total quantity that flows between two specific vertices in a TIN \cite{DBLP:conf/ciac/AkridaCGKS17,DBLP:conf/icde/KosyfakiMPT21}.
Zhou et al. \cite{DBLP:conf/kdd/ZhouZ0H20}
study the problem of dynamically synthesizing realistic TINs by learning from log data.
%Recent studies in cognitive science
%\cite{chen2019resting} associate the information flow in the human brain with the embedded network topology and the interactions between
%different regions. +++
Provenance in TINs can reveal data that can be combined with the structural and flow information of a TIN in order to make pattern detection and graph synthesis more accurate or explain the mining results.

\subsection{Theory and Applications in Provenance}
Buneman et al. \cite{DBLP:conf/icdt/BunemanKT01} were the first who defined and studied the problem of data provenance in database systems.
%from two different perspectives: \textit{why} and \textit{where}.
%According to them, 
The goal of {\em why}-provenance (a.k.a. {\em lineage}) is to explain the existence of a tuple $t$ in a query result by finding the tuples present in the input that contributed to the production of $t$.
%. That is, the reasons behind its inclusion. Why provenance can be answered by finding the path(s) in the query evaluation plan that contribute to the result.
{\em Where}-provenance 
%elaborates more on the relationship between the query input and output. Specifically, for an output tuple $t$, it 
finds the exact attribute values from the input which were copied or transformed to produce $t$.
%Where provenance is much simplier as it only requires to find the tuples in the source tables of the query that contribute to the result.
%Buneman et al. \cite{DBLP:conf/icdt/BunemanKT01} propose a deterministic model for data provenance, 
%where the edges in the query evaluation plan
%are labeled by the data that they use;
%provenance can then be tracked by examining the paths between the query input and output.
An annotation mechanism for where-provenance was proposed in \cite{BunemanKT02}
and implemented in DBNotes
\cite{DBLP:conf/vldb/BhagwatCTV04}.
As query operators (select, project, join, etc.) are executed, annotations are {\em propagated} to eventually reach the query output tuples. 
%This approach assumes queries involve select, project, join, and union only; the where-provenance of an output location is described through a set of propagation rules, one for each relational operator (i.e., select, project, join, union). In contrast to Polygen, these approaches propagate arbitrary annotations through a query, and not only information about source and interme- diary databases.
Annotation propagation depends on the way the query is written/evaluated.
Geerts et al. \cite{DBLP:conf/icde/GeertsKM06} introduced another annotation-based model for the manipulation and querying of both data and provenance,
which
%. Specifically, the model 
allows 
%for the specification of 
annotations on {\em sets of values} and for effectively querying how they are associated. 
%They introduce the concept of {\em block} to represent an annotated set of values. Different colors are given to the blocks representing different annotations. 
The focus is on scientific database curation, where data from possibly multiple sources are integrated and annotations are used to witness the association between the base data that produced a curated tuple.
%Blocks are used to indicate the set of values for which an annotation exists, while block labels are used to indicate the annotations themselves. A new query language is defined which they will refer to as the color algebra. 

Although our solutions share some similarities to provenance approaches for database systems, there are important differences in the data and the propagation models.
First, the TIN graphs that we examine are very large (as opposed to small query graphs) and we track provenance for any vertex in them (i.e., we do not distinguish between input and output vertices). Second, the data transfer model between vertices in TINs is very different compared to data transfer in query graphs. Third, interactions can happen in any order in our TINs, as opposed to query graphs, where edges have a specific order (query graphs are typically DAGs). 
%however in our case we consider (i) the temporal information of the interactions on the edges, (ii) the time of birth of the transferred quantities and (iii) different models for the transfer of quantities between vertices.

Data Provenance  has also been studied in social networks \cite{barbier2013provenance}. 
An important application is to detect 
where from a rumor has started before spreading through the internet. Gundecha et al. \cite{DBLP:conf/cikm/GundechaFL13}
%take into consideration this consensus and try to solve the problem of untraceable information. They
represent social networks as directed graphs and try to recover paths to find out how information spread through the network by isolating important nodes (less than 1\%). The importance of the nodes is based on their centrality.
Taxidou et al. %Another related work in social networks is
\cite{DBLP:conf/www/TaxidouNVFMW15}
studied provenance within an  information diffusion model, based on the W3C Provenance Data Model\footnote{https://www.w3.org/TR/prov-dm/}. A related problem in social networks is information propagation and diffusion. Domingos and Richardson \cite{DBLP:conf/kdd/DomingosR01} were the first who studied techniques for viral marketing to influence social network users. Kempe et al. \cite{DBLP:conf/kdd/KempeKT03} solved the problem of selecting the most influential nodes by proposing a linear model, where the network is represented as a directed graph and vertices are categorized as active and inactive based on their neighbors. 
These approaches are not applicable to TINs, because, in social networks, information {\em is copied and diffused}, whereas in TINs data are {\em moved} (i.e., not copied) from one vertex to another. This key difference makes the provenance problem in TINs unique compared to related problems in previous work.

Savage et
al. \cite{DBLP:conf/sigcomm/SavageWKA00} propose a stochastic packet
marking mechanism that can be used for probabilistic tracing back
packet-flooding attacks in the Internet. The problem setup is quite
different than ours, since we target a more generic provenance problem
in TINs where information is propagated based on various different
models that may not permit backtracing. Moreover, we aim at exact
provenance tracking wherever 
possible.

%Data Provenance has recently gained ground in social networks. Studying the provenance in social networks is vital in order to detect for example, from where a rumor has started before spreading through the internet. Gundecha et al. \cite{DBLP:conf/cikm/GundechaFL13} take into consideration this consensus and try to solve the problem of untraceable information. They represent social networks as directed graphs and try to recover paths to find out how information spread through the network by isolating important nodes (less than 1\%). The importance of the nodes is based on their centrality.
%\todo{ more related work
 % \cite{DBLP:conf/kdd/RichardsonD02},
 % \cite{DBLP:series/synthesis/2013Barbier,DBLP:conf/www/TaxidouNVFMW15},
 % add book 
%}
%These approaches are not applicable to TINs, because, in social networks, information is copied and {\em diffused}, whereas in TINs data are moved from one vertex to another. This key difference makes the provenance problem in TINs unique compared to related problems in previous work.

\subsection{Provenance Systems}

Over the years, a number of systems for provenance tracking have been developed, mainly to serve the need of efficiently storing and managing the annotation data. Chapman et al. \cite{DBLP:conf/sigmod/ChapmanJR08} propose a factorization technique, which 
identifies and unifies common query evaluation subtrees
for reducing the provenance storage requirements.
% and proposepro techniques, based on inheritance and prediction.
Heinis and Alonso \cite{DBLP:conf/sigmod/HeinisA08}
%Another work that addresses the problem of managing efficiently provenance data is \cite{DBLP:conf/sigmod/HeinisA08}.
%In this paper, they
represent workflow provenance mechanisms as DAGs
and compress DAGs with common nodes, in order to save space.
%create intervals to encode the graphs with commons
%nodes in order to eliminate them and reserve more storage.

Several systems \cite{DBLP:conf/vldb/AgrawalBSHNSW06,DBLP:journals/debu/GreenKIT10} have been developed to support the answering of 
{\em data provenance questions}, where the objective is to find how
a data element has appeared in the query result.
% Provenance is considered as a very important concept in databases, especially when we want to track the origin of the information or answer questions related to how a data element was appeared. Many systems have been developed through the years because of the need to record the provenance in order to find answers \cite{DBLP:conf/vldb/AgrawalBSHNSW06} \cite{DBLP:journals/debu/GreenKIT10}.
Karvounarakis et al \cite{DBLP:conf/sigmod/KarvounarakisIT10} developed ProQL, a query language which can be used to detect errors and side effects during the updates of a database. ProQL takes advantage of the graph representation and path expressions to simplify operations involving traversal and projection on the provenance graph. 
%Moreover, they proposed a number of indexing techniques to boost query processing. 
%ProQL can be adapted and used in other systems. 
Titian \cite{DBLP:journals/pvldb/InterlandiSTGYK15} adds provenance support to Spark, aiming at identifying errors during query evaluation.
%of 
%a Spark-based system also tries to identify errors which occur during an execution in databases.

Glavic et al. \cite{GlavicEFT13} present a system for provenance tracking in data stream management systems (DSMS). They
%In this paper, the authors 
propose an {\em operator instrumentation} model, which annotates data tuples that are generated or propagated by the streaming operators with their provenance. They also propose an alternative approach (called {\em replay lazy}), which uses the original operators and, whenever provenance information is needed, the approach replays query processing on the relevant inputs through a instrumented copy of the network (hence, data processing and provenance computation are decoupled). We also propose space-economic models for tracking provenance. However, our input graphs (TINs) are larger and different than DSMS graphs and our propagation models consider the transfer of quantities between vertices as a result of a stream of interactions.

Provenance has also been studied in blockchain systems especially after the huge success of Bitcoin. In \cite {ruan2019fine}, a secure and efficient system called LineageChain is 
implemented on top of Hyperledger\footnote{https://www.hyperledger.org},
for capturing the provenance during contract execution and safely storing it in a Merkle tree.
%LineageChain supports efficient provenance query processing and
%was implemented on top of Hyperledger\footnote{https://www.hyperledger.org}, a well-known framework in blockchain systems.

\section{Definitions}\label{sec:def}
In this section, we formally define the temporal interaction network (TIN)
on which our problem applies.
Then, we present the data propagation model, which determines the
origins of the quantities which are transferred in the network.
Finally, we define the provenance problem that we study in this paper.
Table \ref{table:notations} summarizes the notation used frequently in
the paper.

\begin{definition}[Temporal Interaction Network]\label{def:network}
  A temporal interaction network (TIN) is a directed graph
  $G(V,E,R)$. Each edge
  $(v,u)$ in $E$ captures the history of interactions
  from vertex $v$ to vertex $u$.
  %as a time series $\{(t_1,q_1), (t_2,q_2),\dots\}$.
  $R$ denotes the set of interactions on all edges of $E$.
  Each interaction $r\in R$ is characterized by a quadruple $\langle
  r.s, r.d, r.t, r.q\rangle$, where $r.s\in V$ ($r.d\in V$) is the source
  (destination) vertex of the interaction, $r.t\in {\rm I\!R}^+$ is the time
  when the interaction took place and $r.q\in {\rm I\!R}^+$ is the
  transferred quantity from vertex $r.s$ to $r.d$, due to interaction $r$.
\end{definition}

\begin{figure}[h]
\centering
\subfigure[Interactions]{
\adjustbox{valign=b}{\begin{tabular}{|c|c|c|c|}\hline
$r.s$ & $r.d$ & $r.t$ & $r.q$ \\ \hline
$v_1$ &$v_2$ &1 &3\\	
$v_2$&$v_0$&3&5\\
$v_0$&$v_1$&4&3\\
$v_1$&$v_2$&5&7\\
                       $v_2$&$v_1$&7&2\\
                       $v_2$&$v_0$&8&1\\
                       \hline
        \end{tabular}
  \label{fig:sub2}}}
\subfigure[TIN]{
\adjustbox{valign=b} {   \includegraphics[width=0.35\linewidth]{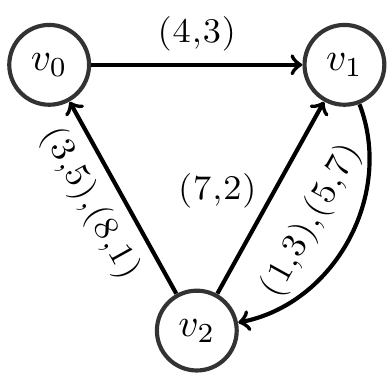}
    \label{fig:sub1}}
}
\caption{A set of interactions and the corresponding TIN}
\label{fig:introex}
\end{figure}

% Fig. \ref{fig:introex}, in the Introduction, shows a sample TIN.
% For example, sequence $\{(3,7),(5,2)\}$ on edge $(v_1,v_2)$
% captures that
% $v_1$ transferred to $v_2$ a quantity of $7$ units at time $3$ and
% then $2$ units at time $5$.
% The corresponding interactions in $R$ are  $\langle v_1,v_2,3,7\rangle$
% and  $\langle v_1,v_2,5,2\rangle$. 

Figure \ref{fig:introex}
%in the Introduction,
shows the set $R$ of
interactions in a TIN and the corresponding graph.
%\todo{move figure here}
For example, sequence $\{(1,3),(5,7)\}$ on edge $(v_1,v_2)$
means that
$v_1$ transferred to $v_2$ a quantity of $3$ units at time $1$ and
then $7$ units at time $5$.
The corresponding interactions in $R$ are  $\langle v_1,v_2,1,3\rangle$
and  $\langle v_1,v_2,5,7\rangle$.

%\makebox[1.3 \textwidth][c]{ 
\begin{table}[ht]
\caption{Table of notations}\label{table:notations}
\vspace{-0.2cm}
\centering
\footnotesize
\begin{tabular}{|c|c|}
\hline
Notation &Description\\ % 
\hline  
%$TIN$ & temporal interaction network\\
%TIN $G(V,E,R)$ & provenance  (problem input)\\
$G(V,E,R)$ & TIN (vertices, edges, interactions)\\
$r.s$ & source vertex of interaction $r\in R$\\
$r.d$ & destination vertex of interaction $r\in R$\\
$r.t$ & time when interaction $r\in R$ took place\\
$r.q$ & transferred quantity during 
interaction $r\in R$\\ 
$B_v$, $|B_v|$ & buffer of vertex $v$, total quantity in $B_v$\\
$O(t,B_v)$ & origin (provenance) data for the quantity at $B_v$ by time $t$\\
$(\tau.o, \tau.q)$ & 
quantity $\tau.q$ originating from $\tau.o$ in $O(t,B_v)$\\
%$\tau.q$ & generated quantity at $\tau.o$ at each buffer $B_v$\\ 
%$resq$ & residue quantity to be transferred for a $r\in R$\\
$\mathbf{p}_v$& provenance vector of a vertex $v\in V$\\  
\hline
\end{tabular}
\end{table}

We consider all interactions $R$ in the TIN {\em in order of time}
and assume that throughout the timeline, each vertex $v\in V$ has a
{\em buffer} $B_v$, which stores the total quantity that has
flown into $v$ but has not been transferred yet to another vertex via
an outgoing interaction from $v$.
We use $|B_v|$ to denote the quantity buffered at $B_v$.

As an effect of an interaction $\langle
  r.s, r.d, r.t, r.q\rangle$, 
vertex $r.s$
transfers a quantity of $r.q$ to vertex $r.d$.
Quantity $r.q$ (or part of it)
could be data that have been accumulated at vertex $r.s$ by
time $r.t$, or $r.q$ could (partially) be generated at $r.s$.
More specifically, we distinguish between two cases:
\begin{itemize}
\item  $|B_{r.s}|\le r.q$. In this case, {\em all} units from $B_{r.s}$ are
  transferred to $B_{r.d}$ due to the interaction.
  In addition, $r.q-|B_{r.s}|$ units are {\em generated} by the source
  vertex $r.s$ and transferred to $B_{r.d}$. Hence, $|B_{r.s}|$
  becomes 0 and $|B_{r.d}|$ is
  increased by $r.q$.
\item  $|B_{r.s}|> r.q$. In this case, $r.q$ units
  are {\em selected} from $B_{r.s}$ to be transferred to $B_{r.d}$.
  Hence, $|B_{r.s}|$ is decreased by $r.q$ and $|B_{r.d}|$ is
  increased by $r.q$. The selection policy may determine the
  routes of the quantities in the network and may affect the result of
  provenance tracking.
\end{itemize}

% $(t,q)$ from vertex $v$ to $u$, $v$
% transfers a quantity of $q$ to $u$.
% The transferred quantity $q$ (or
% part of it) could be data that have been accumulated at vertex $v$ by
% time $t$, or $q$ could (partially) be generated at $v$.
% More specifically, we distinguish between two cases:
% \begin{itemize}
% \item  $|B_v|\le q$. In this case, {\em all} units from $B_v$ are
%   transferred to $B_u$ as a result of interaction $(t,q)$ on edge $(v,u)$.
%   In addition, $q-|B_v|$ units are {\em generated} by $v$ and
%   transferred to $B_u$. Hence, $|B_v|$ becomes 0 and $|B_u|$ is
%   increased by $q$.
% \item  $|B_v|> q$. In this case, $q$ units
%   are {\em selected} from $B_v$ to be transferred to $B_u$.
%   Hence, $|B_v|$ is decreased by $q$ and $|B_u|$ is
%   increased by $q$. The selection policy distinguishes between the
%   different transfer models that we study and affects the result of
%   provenance tracking.
% \end{itemize}

\begin{algorithm}
\begin{algorithmic}[1]
%\LinesNumbered
%\scriptsize
\small
\Require TIN $G(V,E,R)$
\For{each $v\in V$}
      \State $|B_v|=0$ \Comment{Initialize buffers}
\EndFor
\For{each interaction $r\in R$ in order of time}
  \State $q=\min\{r.q,B_{r.s}\}$ \label{lin:transq} \Comment{relayed
    quantity from $B_{r.s}$}
  \State $|B_{r.s}|=|B_{r.s}|-q$\label{lin:decr} \Comment{decrease by $q$}
  \State $|B_{r.d}|=|B_{r.d}|+r.q$\label{lin:incr} \Comment{increase
    by $r.q$, $r.q$$-$$q$ newborn units}
\EndFor
\end{algorithmic}
\caption{Propagation algorithm in a TIN}
\label{algo:greedy}
\end{algorithm}

Algorithm \ref{algo:greedy} is a pseudocode of  the data propagation
procedure. Interactions in $R$ are processed in order of time. For each
interaction $r\in R$, we first determine the {\em relayed} quantity
$q$ from the buffer of the source vertex $r.s$ (Line
\ref{lin:transq}).
This quantity cannot exceed the currently buffered quantity $|B_{r.s}|$
at $r.s$.
Line \ref{lin:decr} decreases $B_{r.s}$, accordingly.
The target node's buffer $B_{r.d}$ is increased by $r.q$ (Line
\ref{lin:incr}).
If $r.q>q$, a new quantity $r.q-q$ is {\em born} by the source
vertex $r.s$ to be transferred to $B_{r.d}$ as part of  $r.q$.

Table \ref{tab:ex_interactions}  shows the
changes in the buffers of the three vertices
in the example TIN (Figure \ref{fig:introex}),
during the application of Algorithm \ref{algo:greedy}.
%set $R$  of interactions
%of the example TIN (Figure \ref{fig:introex})
%ordered by time and the changes in the buffers of the four vertices
%during the application of Algorithm \ref{algo:greedy}.
The values in the parentheses are the newborn quantities at $r.s$,
which are transferred to $r.d$. In the beginning,
all buffers are empty, hence, as a result of the first interaction,
3 quantity units are born at vertex $v_1$ and transferred to $v_2$,
but no previously born quantity is relayed from $B_{v_1}$ to
$B_{v_2}$. At the second interaction, 3 units move from
$B_{v_2}$ to $B_{v_0}$ and 2 {\em newborn units} at $v_2$ are also transferred to  
$B_{v_0}$. At the third interaction, 3 units are {\em selected} to be
transferred from $B_{v_0}$ to
$B_{v_1}$ and no new units are generated because the $B_{v_0}$ had
more
units than $r.q=3$ before the interaction.

% $r.q<|B_{v_1}|$, hence $r.q=2$
% units are {\em selected} from $B_{v_1}$ to be relayed to $B_{v_2}$. 

% \begin{table}[tbh]
%   \caption{Interactions ordered by time}
%   \label{tab:ex_interactions}
%   \vspace{-0.2cm}
%  \centering
%   \footnotesize
%   \begin{tabular}{@{}|c |c |c |c |c |c |c |c |@{}}
%     \hline
%     $r.s$ & $r.d$& $r.t$& $r.q$&$|B_{v_0}|$ & $|B_{v_1}|$& $|B_{v_2}|$& $|B_{v_3}|$\\\hline
%     $v_2$ & $v_3$& $1$& $2$&$0$ & $0$& $0$& $2~(2)$\\\hline
%     $v_0$ & $v_1$& $2$& $5$&$0$ & $5~(5)$& $0$& $2$\\\hline
%     $v_1$ & $v_2$& $3$& $7$&$0$ & $0$& $7~(2)$& $2$\\\hline
%     $v_0$ & $v_1$& $4$& $3$&$0$ & $4~(4)$& $7$& $2$\\\hline
%     $v_1$ & $v_2$& $5$& $2$&$0$ & $2$& $9~(2)$& $2$\\\hline
%    $\dots$&$\dots$&$\dots$&$\dots$&$\dots$&$\dots$&$\dots$&$\dots$\\
%     \hline
%   \end{tabular}
% \end{table}

\begin{table}[tbh]
  \caption{Changes at buffers at each Interaction}
  \label{tab:ex_interactions}
  \vspace{-0.2cm}
 \centering
  \footnotesize
  \begin{tabular}{@{}|c |c |c |c |c |c |c  |@{}}
    \hline
$r.s$ & $r.d$& $r.t$& $r.q$&$|B_{v_0}|$ & $|B_{v_1}|$&
                                                       $|B_{v_2}|$\\\hline
$v_1$&$v_2$&1&3&0&0&3 (3)\\
$v_2$&$v_0$&3&5&5 (2)&0&0\\
$v_0$&$v_1$&4&3&2&3&0\\
$v_1$&$v_2$&5&7&2&0&7 (4)\\
$v_2$&$v_1$&7&2&2&2&5\\
$v_2$&$v_0$&8&1&3&2&4\\
    \hline
  \end{tabular}
\end{table}

Definition \ref{def:provenance}
formally defines the provenance problem that we
study in this paper.
\begin{definition}[Provenance Problem]\label{def:provenance}
  Given a TIN $G(V,E,R)$, at any time moment $t$ and at any vertex $v\in V$
  determine the origin(s) $O(t,B_v)$ of the total quantity accumulated at buffer $B_v$ by
  time $t$. $O(t,B_v)$ is a set of $(\tau.o,\tau.q)$ tuples $\tau$,
  such that each quantity $\tau.q$
  was generated by vertex $\tau.o$ and
  $\sum_{\tau\in O(t,B_v)}\tau.q=|B_v|$.
\end{definition}
At any time $t$, during Algorithm \ref{algo:greedy}, the objective is to
be able to identify the {\em origin} vertices which have generated
the quantities that have been accumulated at buffer $B_v$, for any
vertex $v$. Hence, the problem is to divide the buffer $B_v$ into
a set of $(\tau.o,\tau.q)$ (origin-quantity) pairs, such that each quantity $\tau.q$
is generated by the corresponding vertex $\tau.o$. A data analyst can
then know how the total quantity buffered at $v$ has been composed.

\section{Selection Policies and Provenance}\label{sec:models}
For each interaction $r\in R$,
the selection policy in the case where $|B_{r.s}|> r.q$ determines the
provenance of the quantities that are accumulated at any vertex $v$ (and
transferred from $v$) throughout the timeline.
Selection does make a difference because the quantities in a buffer
$B_{r.s}$ may originate from different vertices.
%Hence, the origins of the quantities in $B_{r.s}$ by time $r.t$ could vary,
%depending of the selection policy.
We present possible selection policies that (i) are based on 
the time quantities are generated, (ii) are based on the order
they are received by the vertex $r.s$ or (iii) choose quantities
proportionally based on their origins. For each policy, we present
annotation mechanisms that can be used to trace the provenance of the
quantities accumulated at the vertices of the TIN. We also discuss
applications where these selection policies may apply.

\subsection{Selection based on generation time}
\label{sec:models:gentime}

The first selection policy is based on the time when the candidate
quantities to be transferred are generated.
% Recall that if $|B_{r.s}|<r.q$,
% a quantity of $r.q-|B_{r.s}|$ units is
% generated by $r.s$ and transferred to $r.d$. This quantity may be relayed
% later to other vertices (as a result of interactions on an outgoing
% edge from $r.d$). At the end of the timeline, we can examine the
% generation origin(s) for each accumulated quantity at each buffer
% $B_v$ to solve the provenance problem for the corresponding vertex $v$.
%The selection policy prioritizes the selection based on the generation
%time of the quantities.
Priority in the selection can be given to the oldest quantities
or the most recently generated ones depending on the
application.
%Without loss of generality,
We will first discuss the {\em
  least recently born} selection policy.  To
implement this approach, any generated quantity should be marked with
the vertex $v$ that generates it and the timestamp $t$ when it is
generated. Hence, during the course of the algorithm, each buffer $B_v$ is
modeled and managed as a set of $(o,t,q)$ triples, where $o$ is the
origin of (i.e., the vertex which bore) quantity $q$ and $t$ is the
time of birth of $q$. The total quantity $|B_v|$ accumulated at buffer
$B_v$ is the sum of all $q$ values in the triples that constitute
$B_v$. As a result of an interaction $r$, if
$|B_{r.s}|>r.q$, the triples in $B_{r.s}$ with the smallest timestamps whose
quantities sum up to $r.q$ are selected and transferred to $B_{r.d}$.
The last triple may be {\em partially transferred} in order for the
transferred quantity to be exactly $r.q$. The triples in each buffer
$B_{v}$ are organized in a min-heap in order to facilitate the
selection.

Algorithm \ref{algo:lru} describes the whole process. For the current
interaction $r\in R$ in order of time, we maintain in variable $resq$
the {\em residue
quantity}, which has yet to be transferred from $r.s$ to $r.d$.
Initially, $resq=r.q$. 
While $q>0$ and $B_{r.s}$ is not empty, we locate the least recently
born triple $\tau$ in $B_{r.s}$ (with the help of the min-heap).
If $\tau.q>q$, this means that
we should transfer part of the quantity in the triple to 
$B_{r.d}$, hence, we {\em split} $\tau$, by keeping it in   
$B_{r.s}$ and reducing $\tau.q$ by $q$ and
initializing a new triple $\tau'$ with the same
origin and birthtime as $\tau$ and quantity $q$.
The new triple is added to $B_{r.d}$. 
If $\tau.q\le q$, we {\em transfer} the entire triple $\tau$ from 
$B_{r.a}$ to $B_{r.d}$.
If $B_{r.s}$ becomes empty and  $resq>0$, then this means that
it was $|B_{r.s}|<r.q$ in the beginning, so we should generate a
newborn triple $\tau'$ with the residue quantity $resq$, having as origin
vertex $r.s$ and marked to be generated at time $r.t$.

\begin{algorithm}
\begin{algorithmic}[1]
%\LinesNumbered
%\scriptsize
\small
\Require TIN $G(V,E,R)$
\For{each $v\in V$}
      \State $B_v=\emptyset$; $|B_v|=0$ \Comment{Initialize buffers}
\EndFor
\For{each interaction $r\in R$ in order of time}
   \State $resq=r.q$ \Comment{residue quantity to be transferred}
   \While {$resq>0$ and $|B_{r.s}|>0$}
      \State $\tau = $ least recent triple in $B_{r.s}$ \Comment{top element
        in heap $B_{r.s}$} \label{lin:lru:selection}
      \If{$\tau.q>resq$} \Comment{split $\tau$}
          \State $\tau'.o=\tau.o$; $\tau'.t=\tau.t$; $\tau'.q=resq$;
          \Comment{new triple}
          \State add  $\tau'$ to $B_{r.d}$; \label{lin:lru:birthsplit}
          \State $\tau.q = \tau.q-r.q$;\Comment{update $\tau$}
          \State $resq=0$; \Comment{transfers completed}
      \Else
         \State remove $\tau$ from $B_{r.s}$ and add it to $B_{r.d}$; \label{lin:lru:transfer}
         \State $resq = resq - \tau.q$ \Comment{update residue quantity} 
      \EndIf
  \EndWhile
  \If{$resq>0$} \Comment{newborn quantity and triple}
     \State $\tau'.o=r.s$; $\tau'.t=r.t$; $\tau'.q=resq$;
     \State add  $\tau'$ to $B_{r.d}$; \label{lin:lru:birth}
  \EndIf
\EndFor
\end{algorithmic}
\caption{Least-recently born selection model}
\label{algo:lru}
\end{algorithm}

Table \ref{tab:oldestfirst} shows the changes in the buffers of the
vertices (shown as sets here, but organized as min-heaps with their
middle element $t$ as key) after each interaction of our running example.
Note that the quantities in the buffers are broken based on their
origins and times of birth.

\begin{table}[tbh]
  \caption{Changes at buffers (oldest-first policy)}
  \label{tab:oldestfirst}
  \vspace{-0.2cm}
 \centering
 \footnotesize
%  \scriptsize
  \begin{tabular}{@{}|c |c |c |c |c |c |c  |@{}}
    \hline
$r.s$ & $r.d$& $r.t$& $r.q$&$B_{v_0}$ & $B_{v_1}$&
                                                       $B_{v_2}$\\\hline
$v_1$&$v_2$&1&3&$\emptyset$&$\emptyset$&\{(1,1,3)\}\\
$v_2$&$v_0$&3&5&\{(1,1,3),(2,3,2)\}&$\emptyset$&$\emptyset$\\
$v_0$&$v_1$&4&3&\{(2,3,2)\}&\{(1,1,3)\}&$\emptyset$\\
$v_1$&$v_2$&5&7&\{(2,3,2)\}&$\emptyset$&\{(1,1,3),(1,5,4)\}\\
$v_2$&$v_1$&7&2&\{(2,3,2)\}&\{(1,1,2)\}&\{(1,1,1),(1,5,4)\}\\
$v_2$&$v_0$&8&1&\{(1,1,1),(2,3,2)\}&\{(1,1,2)\}&\{(1,5,4)\}\\
    \hline
  \end{tabular}
\end{table}

By running
Algorithm \ref{algo:lru},
we can have at any time $t$ the
set of vertices that contribute to a vertex $v$ by time $t$ and the
corresponding quantities (i.e., the solution to Problem \ref{def:provenance}).
In other words, the heap contents for each vertex $v$ at time $t$ corresponds to $O(t,B_v)$.
%This set is in fact the current buffer $B_v$ at time $t$ (i.e., the
%heap contents).
Finally, to implement the {\em most recently born} selection policy, we should
change Line \ref{lin:lru:selection} of Algorithm \ref{algo:lru} to
``$\tau = $ most recent triple in $B_{r.s}$'' and organize each
buffer  as a max-heap (instead of a min-heap).

\stitle{Application}
The least recently born policy is applicable when the
generated quantities lose their value over time (or even expire), which means that the
vertices prefer to keep the most recently generated data. 
On the other hand, the most recently born policy is relevant to
applications, where quantities have {\em antiquity value}, i.e., they
become more valuable as time passes by.

\stitle{Complexity Analysis}
%Algorithm \ref{algo:lru}
%does not incur significant overhead in the
%processing of interactions compared to
%the annotation-free Algorithm \ref{algo:greedy}
%and can help us to obtain at any time the
%vertices that contribute to the
%accumulated quantity at any vertex
In the worst case, each interaction $r$ increases the total number of
triples by one (i.e., by splitting the last transferred triple or by generating a new triple), hence, the space complexity  of the entire process is
$O(|R|)$.
In terms of time, each interaction accesses in the worst case the
entire set of triples at vertex $r.s$. This set is $O(|R|)$ in the worst case, but we expect it to be 
$O(|R|/|V|)$; for each triple in the set, we update two priority queues in
the worst case (i.e, by triple transfers) at an expected cost of $O(\log |R|/|V|)$.
Hence, the overall expected cost (assuming an even distribution of triples) is $O(|R|\cdot |R|/|V| \cdot \log |R|/|V|) = O(|R|^2/|V|\log |R|/|V| )$.

%\todo{add algorithm, add example}

\subsection{Selection based on order  of receipt}
\label{sec:models:receipt}
Another policy would be to select the transferred quantities in order
of their receipt. Specifically, the quantities at each buffer $B_v$
are modeled and managed as a set of $(o,q)$ pairs, where $o$ is the
vertex which generated $q$. These pairs are organized based on the
order by which they have been inserted to $B_v$. If, for the current transaction
$r$, $|B_{r.s}|>r.q$, the last (or the first) quantities in $B_{r.s}$ which sum up
to $r.q$ are selected and added to $B_{r.d}$ in their selection order.
To implement this policy, each buffer is implemented as a FIFO (or
LIFO) queue, hence, it is not necessary to keep track of the
transfer-time timestamps. The algorithm is identical to Algorithm
\ref{algo:lru}, except that Line \ref{lin:lru:selection} becomes 
``least recently added triple in $B_{r.s}$'' in the FIFO policy and
``most recently added triple in $B_{r.s}$'' in the LIFO policy.
Table \ref{tab:lifo} shows the changes in the buffers after each
interaction when the LIFO policy is applied.

\begin{table}[tbh]
  \caption{Changes at buffers (LIFO policy)}
  \label{tab:lifo}
  \vspace{-0.2cm}
 \centering
  \footnotesize
%\scriptsize
%  \begin{tabular}{@{}|@{~}c@{~}|@{~}c@{~}|@{~}c@{~}|@{~}c@{~}|@{~}c@{~}|@{~}c@{~}|@{~}c@{~}|@{}}
  \begin{tabular}{@{}|c|c|c|c|c|c|c|@{}}
    \hline
$r.s$ & $r.d$& $r.t$& $r.q$&$B_{v_0}$ & $B_{v_1}$&
                                                       $B_{v_2}$\\\hline
$v_1$&$v_2$&1&3&$\emptyset$&$\emptyset$&\{(1,3)\}\\
$v_2$&$v_0$&3&5&\{(1,3),(2,2)\}&$\emptyset$&$\emptyset$\\
$v_0$&$v_1$&4&3&\{(1,2)\}&\{(1,1),(2,2)\}&$\emptyset$\\
$v_1$&$v_2$&5&7&\{(1,2)\}&$\emptyset$&\{(1,1),(2,2),(1,4)\}\\
$v_2$&$v_1$&7&2&\{(1,2)\}&\{(1,2)\}&\{(1,1),(2,2),(1,2)\}\\
$v_2$&$v_0$&8&1&\{(1,2),(1,1)\}&\{(1,2)\}&\{(1,1),(2,2),(1,1)\}\\
    \hline
  \end{tabular}
\end{table}

\stitle{Application}
The FIFO policy is used in applications where the buffers
are naturally implemented as FIFO queues (pipelines, traffic
networks).
The LIFO policy applies when the accumulated quantities
are organized in a stack
(e.g., cash registers, wallets) before being transferred.

\stitle{Complexity Analysis}
The space complexity is $O(|R|)$, same as that of 
generation time selection policies (Sec. \ref{sec:models:gentime}),
because the only change is that we replace the heap by a FIFO queue (or
a stack).
This replacement changes the
access and update costs from
$O(\log |R|/|V|)$ to $O(1)$.
%The time complexity can also be
%except that the replacement of the heap data structure at each vertex
%by a FIFO queue or a stack changes the access and update costs from
%$O(\log |R|/|V|)$ to $O(1)$.
Hence, the overall expected cost is reduced from \linebreak $O(|R|^2/|V|\log |R|/|V| )$ to $O(|R|^2/|V|)$.

\subsection{Proportional selection}
\label{sec:models:proportional}

The proportional selection policy,
for the case where $|B_{r.s}|> r.q$,
chooses the relayed quantity from $r.s$ to $r.d$ {\em proportionally} from the vertices that
have contributed to $B_{r.s}$, based on their
contribution.

Formally, for each vertex $v\in V$, we  define a $|V|$-length
vector $\mathbf{p}_v$, which captures the provenance of the
quantity currently in its buffer $B_v$.
The $i$-th value of $\mathbf{p}_v$ is the quantity fragment in $B_v$ which originates
from the $i$-th vertex of the TIN $G$.
Hence, the sum of quantities in $\mathbf{p}_v$ equals the total quantity
$|B_v|$ in $B_v$. Initially, all values of $\mathbf{p}_v$ are 0.

Algorithm \ref{algo:proportional} 
shows how the provenance vectors are updated after each interaction $r$.
%presents the steps of an algorithm
%which computes the accumulated quantities at each vertex as a result
%of each interaction, while updating the affected provenance vectors.
We distinguish between two cases. The first one is when  $r.q \ge
|B_{r.s}|$, i.e., the quantity $r.q$ to be transferred by the current
interaction is greater than or equal to the buffered quantity
$|B_{r.s}|$ at the source buffer. In this case, the entire buffered
quantity in  $B_{r.s}$ is relayed to $B_{r.d}$.
Hence, vector $\mathbf{p}_{r.s}$ is added to $\mathbf{p}_{r.d}$
(symbol $\oplus$ denotes vector-wise addition).
If $r.q$ is strictly greater than $|B_{r.s}|$, a newborn quantity $r.q
- |B_{r.s}|$ at $r.s$ is added to $B_{r.d}$, hence, we should add the
corresponding provenance information to the $r.s$-th element of
$\mathbf{p}_{r.d}$ (Line \ref{lin:prop:newborn}).
 This is denoted by the addition of vector
 $\mathbf{e}_{r.s,(r.q-B_{r.s})}$, where $\mathbf{e}_{v,x}$ denotes a
 vector with all $0$'s except having value $x$ at position $v$. 
The second case is when $r.q < |B_{r.s}|$. In this case, the quantity
$r.q$ which is transferred from $r.s$ to $r.d$ is chosen
proportionally. Specifically, if vertex $r.s$ has in its
buffer $B_{r.s}$ a quantity $q$ which was born by the $i$-th vertex, then a
quantity $q\cdot \frac{r.q}{|B_{r.s}|}$ should be transferred from
the $i$-th position of $\mathbf{p}_{r.s}$ to the 
$i$-th position of $\mathbf{p}_{r.d}$.
This translates into the vector-wise operations at Lines \ref{lin:prop:oplus}
and \ref{lin:prop:ominus} of Algorithm \ref{algo:proportional}.
Table \ref{tab:propo} shows the changes in the buffer vectors after each
interaction when proportional selection is applied.

\begin{algorithm}
\begin{algorithmic}[1]
%\LinesNumbered
%\scriptsize
\small
\Require TIN $G(V,E,R)$
\For{each $v\in V$}
      \State $|B_v|=0$; $\mathbf{p}_v=\mathbf{0}$; \Comment{Initialize buffers
        and vectors}
\EndFor
\For{each interaction $r\in R$ in order of time}
  \If{$r.q \ge |B_{r.s}|$}
  \State $\mathbf{p}_{r.d}  = \mathbf{p}_{r.d} \oplus \mathbf{p}_{r.s} \oplus
  \mathbf{e}_{r.s,(r.q-B_{r.s})}$; $\mathbf{p}_{r.s}=\mathbf{0}$; \label{lin:prop:newborn}
  \State $|B_{r.d}|=|B_{r.d}|+r.q$; $|B_{r.s}|=0$; 
  \Else \Comment{$r.q < |B_{r.s}|$}
  \State $\mathbf{p}_{r.d}  = \mathbf{p}_{r.d} \oplus (r.q/|B_{r.s}|) \mathbf{p}_{r.s}$; $B_{r.d}=B_{r.d}+r.q$; \label{lin:prop:oplus}
  \State $\mathbf{p}_{r.s}  = \mathbf{p}_{r.s} \ominus (r.q/|B_{r.s}|) \mathbf{p}_{r.s}$; $B_{r.s}=B_{r.s}-r.q$; \label{lin:prop:ominus}
  \EndIf
\EndFor
\end{algorithmic}
\caption{Proportional selection model}
\label{algo:proportional}
\end{algorithm}

\begin{table}[tbh]
  \caption{Changes at buffers (proportional selection)}
  \label{tab:propo}
  \vspace{-0.2cm}
 \centering
  \footnotesize
  \begin{tabular}{@{}|c |c |c |c |c |c |c  |@{}}
    \hline
$r.s$ & $r.d$& $r.t$& $r.q$&$\mathbf{p}_{v_0}$ & $\mathbf{p}_{v_1}$&
                                                       $\mathbf{p}_{v_2}$\\\hline
$v_1$&$v_2$&1&3&$[0,0,0]$&$[0,0,0]$&$[0,3,0]$\\
$v_2$&$v_0$&3&5&$[0,3,2]$&$[0,0,0]$&$[0,0,0]$\\
$v_0$&$v_1$&4&3&$[0,1.2,0.8]$&$[0,1.8,1.2]$&$[0,0,0]$\\
$v_1$&$v_2$&5&7&$[0,1.2,0.8]$&$[0,0,0]$&$[0,5.8,1.2]$\\
$v_2$&$v_1$&7&2&$[0,1.2,0.8]$&$[0,1.66,0.34]$&$[0,4.14,0.86]$\\
$v_2$&$v_0$&8&1&$[0,2.03,0.97]$&$[0,1.66,0.34]$&$[0,3.31,0.69]$\\
    \hline
  \end{tabular}
\end{table}

\stitle{Application}
Proportional selection makes sense in applications where the
quantities are naturally mixed in the buffers.
This includes cases when the transferred data are liquids (e.g.,
buffers are oil tanks) or indistinguishable financial units in accounts
(i.e., balances in bank accounts, capital stocks in digital
portfolios). In such cases, it is reasonable to consider that the
origins of the buffered quantities contribute proportionally to a transfer.

\stitle{Complexity Analysis}
The provenance vectors $\mathbf{p}_v$ raise the space requirements of
this model to $O(|V|^2)$, i.e., we need a $|V|$-length vector for
each vertex.
In the next section, we will explore a number of directions in order
to reduce the space requirements and make
proportional provenance tracking feasible for large graphs with
millions of vertices.
The time complexity is also high, because we need
one or two vector-wise operations per interaction, which accumulates
to a $O(|R|\cdot |V|)$ cost.
In our implementation, we exploit SIMD instructions \cite{polychroniou2015rethinking} to reduce the cost of vector-wise operations.
%If $|V|$ is not very large, this cost is reduced in practice by 

%\subsection{Using sparse vector representations}
%\label{sec:opt:sparse}
%The proportional model requires the maintenance of a $|V|$-length
%vector for each vertex $v$ of the graph. 
\stitle{Sparse vector representations}
In sparse graphs,
each vertex $v$ may receive quantities originating from a small subset of vertices
in practice. 
To save space, instead of storing each space-demanding vector
$\mathbf{p}_v$ explicitly, we can represent it by
an ordered list of $(u, q)$ pairs, for each vertex $u$ contributing a
quantity $q>0$ in the buffer $B_v$.
For example, after the temporally
first interaction in our running example,
instead of storing $\mathbf{p}_{v_2}$
as $[0,3,0]$, we store it as $[(v_1,3)]$, implying that $v_2$
received its 3 units from $v_1$.
The vector update operations of Algorithm \ref{algo:proportional} can be replaced by merging the ordered lists of the corresponding sparse vector representations. 
This way, the space requirements are reduced from
$O(|V|^2|)$ to  $O(|V|\cdot \ell)$, where $\ell$ is the average length of
the list representations of the vectors. The time complexity is
reduced to $O(|R|\cdot \ell)$, accordingly. Still, as we show
experimentally, in Section \ref{sec:exps}, $\ell$ can grow too large
and we may not be able to accommodate the lists in memory, after a
long sequence of interactions.
%The reason for this is that for each interaction $r$, where $r .q <
%|B_{r .s}|$ (i.e., selection is applied), the number of entries that
%are added to $\mathbf{p}_{r.d}$ can be as many as the entries in
%$\mathbf{p}_{r.s}$, which means that the $i$-th interaction $r$ can add as
%many as $\min\{i,|V|\}$ interactions to $\mathbf{p}_{r.d}$.
%In the rest of the section, we present
%techniques that address the memory issue, by limiting the scope of provenance tracking.

%\input{problem}
\section{Scalable Proportional Provenance}\label{sec:algorithms}
Proportional provenance tracking
(Section
\ref{sec:models:proportional})
has high space and time complexity compared to the models based on generation time (Section
\ref{sec:models:gentime}) or receipt order (Section
\ref{sec:models:receipt}).
%; hence, this approach may not be applicable on large graphs.
%
%%In this section, we identify the computational bottlenecks of the provenance tracking
%%models of Section \ref{sec:models} and propose solutions for them.
%Tracking provenance using the models based on generation time (Section
%\ref{sec:models:gentime}) or receipt order (Section
%\ref{sec:models:receipt})
%has relatively low space and time complexity
%and does not incur significant overhead compared simply running
%Algorithm \ref{algo:greedy}, which does not include any provenance
%tracking mechanism.
%On the other hand, the proportional selection model (Section
%\ref{sec:models:proportional}) has high space complexity and
%update cost, which may
%render this approach inapplicable on large graphs.
We investigate a number of techniques that reduce
the space requirements and constitute
proportional provenance feasible even on very large graphs.

\subsection{Selective provenance tracking}
\label{sec:opt:selective}
In many applications, we may not have to track provenance from all
vertices in the graph, but from a selected subset thereof. For
example, in a financial network, we could limit our focus to a
specific set of entities, suspected to be involved in
illegal activities.
Or, we may select the $k$
vertices that generate the largest total quantities.
% contributing vertices (these can be
%easily found by changing Line \ref{lin:incr} of
%Algorithm \ref{algo:greedy} to measure the
%total quantity generated by each vertex).
Hence, we can limit the size of the provenance
tracking vectors $\mathbf{p}_v$ to include only a given subset of
vertices having limited size $k$. %However, we cannot exclude from the
%process the remaining $|V|-k$ vertices because they should be
%considered in the proportional transfer of quantities at each
%interaction.

Specifically, for each vertex $v\in V$, we maintain a vector $\mathbf{p}_v$ of size $k+1$, where the first
$k$ positions correspond to the vertices of interest and the last
position represents the rest of the vertices.
Algorithm \ref{algo:proportional} can now directly be applied, after
the following change: if any of the source vertex $r.s$ or the destination
$r.d$ is not in the set of the $k$ vertices of interest, we update
the $(k+1)$-th position, which accumulates the sum of quantities from
all vertices except the selected ones.
This version of proportional selection algorithm has reduced space and
time complexity compared to Algorithm
\ref{algo:proportional}. Specifically, its space requirements are
$O(k\cdot |V|)$ and its time complexity is
$O(k\cdot |R|)$.

\subsection{Grouped provenance tracking}
\label{sec:opt:groups}
In practice, tracking provenance from individual vertices could
provide too many details which might be hard to interpret.
It might be more practical, to track provenance from groups of
vertices. Hence, assuming that the vertices of the TIN have been
divided into groups, we can replace the long $\mathbf{p}_v$ vectors by
shorter vectors of length $m$, where $m$ is the number of groups. This
means that, in the end, for each vertex $v$ we will have in
$\mathbf{p}_v$ the total quantity in the buffer $B_v$ of $v$
which originates from each group. The grouping of vertices can be done
in different  ways depending on the application. For example, the
values of one or more attributes that characterize the
vertices in the application (e.g., gender, country) can be used for
grouping. Network clustering algorithms (e.g., METIS \cite{DBLP:journals/siamsc/KarypisK98}) or geographical
clustering can be used to divide the vertices to groups.

Algorithm \ref{algo:proportional} can easily be adapted to operate on
groups. The vertices involved at each interaction (i.e., $r.s$ and
$r.d$)
are mapped to group-ids and the
corresponding positions are updated in the vector-wise operations.
% Specifically, each vertex $v\in V$ is {\em mapped} to a group $mf(v)$. 
% In Algorithm \ref{algo:proportional}, the vector 
% $\mathbf{p}_{v}$ of each $v\in V$ has length $k$, where $k$ is the
% number of groups and, at line 6 of Algorithm \ref{algo:proportional},
% $\mathbf{e}_{mf(r.s),(r.q-B_{r.s})}$ is used to update 
% are conducted after converting the
% origin.
% Then, 
As in the case of selective provenance tracking (Section
\ref{sec:opt:selective}), the space and time complexity is reduced to
$O(m \cdot |V|)$ and 
$O(m \cdot |R|)$, respectively.

% \subsection{Data parallelism}
% \label{sec:opt:simd}

% \fix{discussion to be moved to section where we describe proportional provenance}

% Since Algorithm \ref{algo:proportional} involves vector-wise
% operations, we can exploit SIMD instructions to improve its
% efficiency.
% Specifically, +++

% \subsection{Streaming data and Sliding Windows}\label{sec:streaming}
\subsection{Limiting the scope of provenance}\label{sec:streaming}
If selective and grouped provenance is not an option, tracking proportional provenance in 
large graphs with millions of vertices could be infeasible.
We investigate two techniques that limit the scope of provenance
by either avoiding the tracking of quantities generated far in the past or setting a budget for provenance at each vertex.
Our techniques are especially suitable for streaming data, where 
speed and feasibility are preferred over preciseness.

\subsubsection{Windowing approach}
Our first approach takes as input a parameter $W$, representing a {\em
  window}, which determines how far in the past we are interested in
tracking provenance.  
Specifically, for each vertex $v$ we can guarantee finding the provenance of quantities
that reach $v$, which where born up to $W$ interactions before.
To achieve this, for each $v$, we initialize {\em two} sparse
(i.e., list) provenance vector representations  
$\mathbf{p}^{odd}_v$ and
$\mathbf{p}^{even}_v$.
At each interaction, both lists are updated. 
%as described in Section
%\ref{sec:models}. 
However, whenever we reach an interaction $r$ whose
order is a multiple of $W$, we {\em reset}  either
$\mathbf{p}^{odd}_v$ or
$\mathbf{p}^{even}_v$ as follows.
If the order of $r$ in the sequence $R$ of interactions
is an odd multiple of $W$, for each vertex $v\in V$, we reset its provenance
  list $\mathbf{p}^{odd}_v$ by setting $\mathbf{p}^{odd}_v=[(\alpha, |B_v|)]$,
  where $\alpha$ is an {\em artificial vertex}, representing the
  entire set $V$ of vertices.
  This means that we assume that the entire quantity in $B_v$
  has unknown provenance.
  If the order of $r$
  is an even multiple of $W$, for all vertices $v$, we reset $\mathbf{p}^{even}_v$ by setting $\mathbf{p}^{even}_v=[(\alpha, |B_v|)]$.
%With the mechanism described above, we achieve the following. First,
After any interaction $r$, we can track provenance for any vertex $v$
using whichever of $\mathbf{p}^{even}_v$ or $\mathbf{p}^{odd}_v$ was
least recently reset. This guarantees that we can track the provenance
of quantities born at least $W$ (and at most $2\cdot W$) interactions
before.
The space requirements (i.e., the total space required to
store the provenance lists) are now controlled due to the provenance list
resets.

\begin{figure}[h]
\centering
\includegraphics[width=0.8\linewidth]{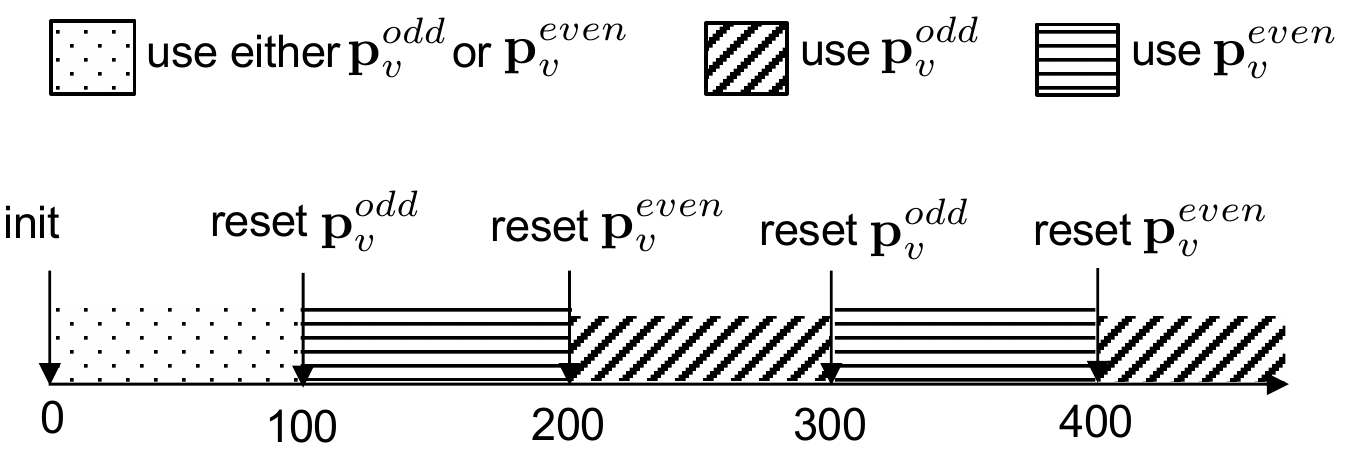}
\caption{Windowing approach in provenance tracking}
\label{fig:doublebuffer}
\end{figure}

Figure \ref{fig:doublebuffer} illustrates how, for each vertex $v$,
$\mathbf{p}^{odd}_v$ and $\mathbf{p}^{even}_v$ are updated and used.
Assuming that $W=100$, until the $100$-th interaction,
$\mathbf{p}^{odd}_v$ and $\mathbf{p}^{even}_v$ are identical and
either of them can be used. Since $\mathbf{p}^{odd}_v$  is reset at
the $100$-th interaction, between the $100$-th and the $200$-th
interaction   $\mathbf{p}^{even}_v$ is used to track the provenance of
quantities which were generated since the first interaction.
Similarly, between the $200$-th and the $300$-th
interaction   $\mathbf{p}^{odd}_v$ is used to track provenance up to
the $100$-th interaction.

\subsubsection{Budget-based provenance}

Another approach which we can apply to control the memory requirements
and make proportional provenance tracking feasible on large graphs is
to allocate a maximum capacity $C$ (budget) to each vertex $v$ for its
provenance list $\mathbf{p}_v$.
Whenever we have to add new entries to $\mathbf{p}_v$, if
the required capacity after the addition exceeds $C$, we
select a certain fraction $f$ of entries to keep in
$\mathbf{p}_v$. We remove the remaining entries and
assume that the total
quantity $Q$ which originates from them
was born at an artificial vertex $\alpha$, modeling all vertices
(i.e., unknown source).
Hence, if $\mathbf{p}_v$ includes an $(\alpha,q)$ entry, the entry
is updated to $(\alpha,q+Q)$; if not, a new entry $(\alpha,Q)$ is added to
$\mathbf{p}_v$.

With this approach, the space requirements of proportional provenance
tracking become $O(|V|\cdot C)$. The larger the value of $C$ the
more accurate provenance tracking becomes. Parameter $f$ should be
chosen such that the memory allocated at each vertex is not
underutilized and, at the same time,
shrinking does not happen very often. We suggest a value
between $0.6$ and $0.8$.
Finally, the selection of entries to keep when the budget $C$ is
reached in $\mathbf{p}_v$
can be done using different
criteria. For example, we can keep the entries with the largest
quantities, or set a priority/importance order to vertices and keep 
provenance data for the most important ones.

As an example, assume that $\mathbf{p}_v =
\{(v,1),(u,3),(w,2),(z,1)\}$ and let $C=5$.
Let $\{(x,2),(w,1),(y,4)\}$ be the new entries that have to be added/merged
into $\mathbf{p}_v$.
After the change, $\mathbf{p}_v$ should become $\mathbf{p}_v =
\{(v,1),(u,3),(w,3),(x,2),(y,4),(z,1)\}$, i.e., the capacity
constraint $C=5$ is violated.
If $f=0.6$, we should keep $0.6\cdot C=3$ entries; let us assume that
we keep the ones with the largest quantities, i.e.,
$\{(u,3),(w,3),(y,4)\}$. The remaining three entries are replaced in $\mathbf{p}_v$
by an entry $(\alpha,4)$, since the sum of their quantities is
4. Hence,  after the update,  $\mathbf{p}_v$ becomes $\{(u,3),(w,3),(y,4),(\alpha,4)\}$.
Note that selecting the entries with the largest quantities may cause a bias in favor of origins that generate quantities early over origins whose generation is spread more evenly in the timeline.

\section{Tracking the paths}\label{sec:howprovenance}
So far, we have studied the problem of identifying the origins of the
quantities accumulated at the vertices.
%This corresponds to a {\em where}-provenance problem in query
%evaluation \cite{DBLP:journals/ftdb/CheneyCT09}.
%While this could be enough in many
%applications, 
An additional
question is which path did each of the quantities, accumulated at a
vertex $v$, follow from its origin to $v$.
This information can provide more detailed {\em explanation} for the reasons behind
data transfers and corresponds to {\em how}-provenance in query
evaluation \cite{DBLP:journals/ftdb/CheneyCT09}.

To implement how-provenance
for the selection models of Sections \ref{sec:models:gentime}
and \ref{sec:models:receipt},
for each quantity element in the buffer $B_v$ of
every node $v$, we maintain
a {\em transfer} path, which
captures the route that the element has followed so far from its
origin to $v$.
When a new quantity element is generated, either as a result
of a {\em split} (i.e., Line \ref{lin:lru:birthsplit} of Algorithm
\ref{algo:lru}),
or anew  (i.e., Line \ref{lin:lru:birth} of Algorithm \ref{algo:lru}),
its path is initialized to include just the origin vertex $r.s$.
%the edge $(r.s,r.d)$.
Every time a quantity element is transferred from one
vertex to another as a result of an interaction $r'$ (i.e., Line \ref{lin:lru:transfer} of Algorithm
\ref{algo:lru}), its path is extended to include the transmitter
vertex $r'.s$.
%by edge $(r.s,r.d)$.
This way, for each quantity element, we keep track of not just its
origin but also the path which the quantity has followed. 

Note that path tracking in the case of proportional selection is
not meaningful, because, if $r.q < |B_{r.s}|$,  all quantities in
$B_{r.s}$ are split to a fraction that remains at $B_{r.s}$ and a
fraction that moves to $B_{r.s}$, wherein they are {\em combined} with
the corresponding quantities from the same origins. This means that 
quantities in a buffer from the same origin (but potentially
from multiple different paths) are mixed and indistinguishable.  

\stitle{Complexity Analysis}
Path tracking does not change the time complexity, as the number of
path changes is $O(|R|)$ and each path initialization or extension costs $O(1)$.
On the other hand, the space complexity increases by a factor of
$O(|R|/|V|)$, i.e., the expected number of quantity element transfers
(executions of Line \ref{lin:lru:transfer} of Algorithm \ref{algo:lru}).
Hence, the time complexity increases to $O(|R|^2/|V|)$.

\section{Experimental Evaluation}\label{sec:exps}
In this section, we experimentally evaluate the performance and
scalability of our proposed provenance tracking techniques, which
apply to the different selection models presented in Section
\ref{sec:models}. For this, we used five real TINs, described in
Section \ref{sec:datasets}.
We compare the different selection policies for information
propagation in terms of runtime cost and memory requirements in
Section \ref{sec:exp:perf}.
In Section \ref{sec:exp:sel}, we evaluate the performance of selective
and grouped provenance tracking using the proportional selection
policy.
Section \ref{sec:exp:scope} tests the windowing and budget-based
approaches for limiting the scope of provenance tracking.
Section \ref{sec:exp:paths} evaluates the memory and
computational overheads of tracking the paths of quantities
accumulated at each vertex.
Finally, Section \ref{sec:exp:usecase} presents a use case that demonstrates the practicality of provenance in TINs .
All provenance tracking methods were implemented in C and compiled
using gcc with -O3 flag.
The experiments were run on a machine with a 3.6GHz Intel i9-10850k
processor and 32GB RAM.  
%and \fix{study the differences among
%  the models}.
%We implemented the algorithms presented in Section 4
%and section 5 respectively.

\subsection{Description of datasets}\label{sec:datasets}
%We used five datasets extracted from real networks: Bitcoin Network,
%IP addresses, Prosper Loans, Flights Network and Taxis Network.
Table
\ref{table:datasets} summarizes the statistics for each of the
datasets that we use in the experiments. Below, we provide a detailed
description for each of them. 

\stitle{Bitcoin Network:}
This dataset includes all transactions in the bitcoin network up
to 2013.12.28; we considered these transactions as interactions.
The data were preprocessed and made available by the authors
of \cite{DBLP:journals/corr/KondorPCV13}. We merged
bitcoin addresses which belong to the same user.
For each interaction, as quantity, we consider the corresponding amount of BTCs
exchanged between the addresses.
We converted all amounts to BTC (originally
Satoshi) and we did not take into consideration transactions with
insignificant flow (i.e., less than 0.0001 BTC). Data provenance in
this network can unveil the funding sources of addresses and
explain the reasons behind bitcoin exchanges.

\stitle{CTU Network:} A Botnet traffic network was extracted and
created by CTU University
\cite{DBLP:journals/compsec/GarciaGSZ14}. We used the data and we
designed a TIN. The vertices are the IP addresses and
the interactions are the transactions among the nodes at different time
periods. The quantity of each interaction is the total amount of
bytes which are transferred between the corresponding vertices.
Tracing the provenance of quantities that reach vertices in such a
network may help toward analysis of potential network attacks.
%to  
%problem in such networks is very useful especially when for
%example an attack happens and it is crucial to answer questions
%related to the origin of the attack.

\stitle{Prosper Loans:} We downloaded this
dataset from http://konect.cc and
created the corresponding interaction
network. The vertices of the network correspond to users and the
interactions represent loans between them.
The lended amounts are the quantities that are exchanged at the
interactions.
Tracking the provenance of amounts that reach certain nodes may help
in the identification of the direct or indirect relationships between lenders and borrowers. 
%The structure of the Prosper Loans
%network has some similarities with the Bitcoin Network because of the
%transactions which take place between the nodes. Users lend money to
%other users. The vertices of the graph are the lenders and the
%borrowers respectively and the edges are the transactions which
%carries a quantity (the loan amount) between them at
%different timestamp. 

\stitle{Flights Network:} We extracted flights data from Kaggle%
\footnote{https://www.kaggle.com/yuanyuwendymu/airline-delay-and-cancellation-data-2009-2018}.
We converted the original file into an interaction network, where
vertices are the origin and destination airports and the time of
departure was used to model the time of the corresponding
interaction.
We used the number of passengers in each flight as the 
quantity in the corresponding interaction.
Since this number was not given in the original data, we
have put a random number between 50 and 200.
Provenance information can help us understand the reasons behind
potential traffic, bottlenecks, or other issues at airports. 

\stitle{Taxis Network:}
We considered NYC yellow taxi trips\footnote{https://www1.nyc.gov/site/tlc/about/tlc-trip-record-data.page}
on January 1st 2019 as interactions in a TIN, where vertices are
taxi zones (pick-up and drop-off districts), the drop-off time
represents the time of interactions and the number of passengers are
the corresponding quantities.
Similar to the flights network, we can apply provenance tracking to
investigate the reasons behind the accumulation of passengers at
different zones.
%information can help us understand the reasons behind
%potential traffic or other issues at airports. 

\begin{table}[ht]
\caption{Characteristics of Datasets}
\vspace{-0.2cm}
\centering
\small
 %\scriptsize
\begin{tabular}{|l|c|c|c|}
\hline
Dataset &\#nodes & \#interactions &average $r.q$ \\
\hline
Bitcoin &12M&45.5M & 34.4\bitcoinB\\
CTU&608K&2.8M& 19.2KB\\
Prosper Loans &100K&3.08M&$\$$76\\
Flights &629&5.7M&125\\  
 Taxis &255&231K&1.53\\[0.2ex]
\hline
\end{tabular}
\label{table:datasets}
\end{table}

\subsection{Provenance tracking performance}\label{sec:exp:perf}
In our first set of experiments, we investigate the runtime cost and
the memory requirements of provenance tracking based on the different
selection policies for information propagation, presented in Section
\ref{sec:models}.
We executed each method by processing the entire sequence of
interactions and updating the necessary information for each of them,
according to the algorithms described in Section \ref{sec:models}. 
Tables \ref{table:runtimeall} and
\ref{table:memoryall} show the runtime cost and the peak memory use by
the different selection policies.
As a point of reference we also included the basic propagation
algorithm that does not track provenance (Algorithm
\ref{algo:greedy}), denoted by NoProv.

From the two tables, we observe that
the methods based on generation time (Section  \ref{sec:models:gentime})
are scaleable, since they terminate even at very large graphs with
millions of interactions (i.e., Bitcoin network). Naturally, they are
one to two orders of magnitude slower than NoProv, as NoProv has $O(1)$ cost per interaction.
  Their space overhead compared to NoProv is not
  high for big and sparse graphs, like Bitcoin and CTU. On the
  other hand, for smaller graphs with heavy traffic between vertices,
  the space requirements become high. 
%In terms of space, \fix{write after verification of mem. req.}

  The methods that select the information to propagate
  based on order of receipt (Section  \ref{sec:models:receipt})
  are also slower than NoProv, but faster than the ones that use
  generation time, because they do not have to maintain a heap and
  select the propagated quantities from it. Instead, the simpler data
  structures that they use (stack, FIFO queue) are more efficient. 
  In terms of space, their requirements are lower compared to the
  order-of-receipt policies mainly because they do not need to store
  and propagate the time of birth together with the origin vertices
  (i.e., each provenance tuple has two values instead of three).
  Their behavior in big/sparse graphs compared to small/dense ones is
  similar to the one of order-of-receipt policies discussed above.
%  \fix{write after verification of mem. req.}

  As opposed to the selection policies of Sections
  \ref{sec:models:gentime} and \ref{sec:models:receipt}, the
  proportional selection policy, presented in Section
  \ref{sec:models:proportional},
  performs best when the number of
  vertices in the graph is small (i.e., at the Flights and Taxis
  networks). This is expected because their storage overhead in this
  case is manageable (at most $O(|V|^2)$). Specifically, the
  proportional policy 
  using dense vector representations
%  described in Section
%  \ref{sec:models:proportional}, denoted by proportional ({\em dense})
  can be used only for the Flights and Taxis
  networks, with very good performance.
    Even when the sparse
  vector representations are used,
  the required memory exceeds the capacity of our machine in the
  Bitcoin and CTU networks. This approach can be used on the
  Prosper Loans network, however, it requires a lot of space (2.4GB)
  and it is significantly slower than the  policies  of Sections
  \ref{sec:models:gentime} and \ref{sec:models:receipt}, because it
  needs to manage and maintain long lists.
  This necessitates the use of the scope limiting techniques described
  in Section \ref{sec:streaming}, as tracking provenance from all
  vertices in the entire
  history of interactions becomes infeasible.
  %of selective/grouped provenance tracking ++

\begin{table*}
  \small
  \caption{Runtime (sec) for each selection policy}
  \label{table:runtimeall}
  \centering
  \vspace{-0.3cm}
  \begin{tabular}{|l|c|c|c|c|c|c|c|}
    \hline
    Dataset&
    No Provenance&
    Least Recently Born&
    Most Recently Born&
    LIFO&
    FIFO&
          Proportional (dense)&
                                Proportional (sparse)\\
    \hline

    Bitcoin&0.19&31.77 &9.17 &3.10 &3.90&--&--\\
    CTU&0.010&0.16 &0.19 & 0.08&0.11&--&--\\
    Prosper Loans&0.006& 0.089&0.082&0.055 &0.08&--&15.7\\
    Flights&0.009&0.75 &0.77 &0.077 &0.15&1.58&2.91\\
    Taxis&0.0005&0.014 &0.015 &0.002 &0.004&0.032&0.05\\
    %\hline \!\!
    
    \hline
  \end{tabular}
\end{table*}

\begin{table*}
  \small
  \caption{Peak memory used by each selection policy}
  \label{table:memoryall}
  \centering
  \vspace{-0.3cm}
  \begin{tabular}{|l|c|c|c|c|c|c|c|}
    \hline
    Dataset&
    No Provenance&
    Least Recently Born&
    Most Recently Born&
    LIFO&
    FIFO&
          Proportional (dense)&
                                Proportional (sparse)\\
    \hline
    Bitcoin&96MB&891MB & 892MB&536MB &535MB&--&--\\
    CTU&4.85MB& 56.4MB&56.4MB &33.8MB &33.8MB&--&--\\
    Prosper Loans&800KB&61.4MB &61.4MB&36.8MB &36.8MB&--&2.4GB\\
    Flights&5KB&0.90MB&1.05MB&1.05MB&1.05MB &3.16MB&2.32MB\\
    Taxis&2KB&0.93MB &1.02MB &0.59MB&0.6MB&0.52MB&0.44MB\\
    %\hline \!\!
    
    \hline
  \end{tabular}
\end{table*}

\subsection{Selective and grouped provenance }\label{sec:exp:sel}
In the next set of experiments, we evaluate the performance of
proportional provenance only for a subset of vertices or for groups of
vertices as described in Sections \ref{sec:opt:selective} and
\ref{sec:opt:groups}.
We conduct the experiments on the three largest networks (in terms of
number of vertices), i.e., Bitcoin, CTU, and Prosper Loans. Recall
that on these networks tracking proportional provenance from all
vertices is infeasible or very expensive. Let $k$ denote the
number of selected vertices (for selective provenance) or the number
of groups (for grouped provenance). We measure the runtime cost and
memory requirements of proportional provenance for different values of
$k$. In the case of  selective provenance, we select the top-$k$
contributing vertices as the set of vertices for which we will measure
provenance. That is, we first run  NoProv (Algorithm
\ref{algo:greedy}) and measure the total quantity generated by each vertex
and then choose the ones that generate the largest quantity.
In case of grouped provenance, we randomly allocate vertices to groups
in a round-robin fashion; since the runtime performance and memory
requirements are not affected by the group sizes or the way the
vertices are allocated to groups, this allocation does not affect the
experimental results.

\begin{figure*}[t!]
 \includegraphics[width=0.60\textwidth]{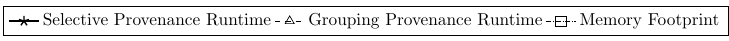}\\
\subfigure[Bitcoin Network]{
   \label{fig:exp:bitcoin}
    \includegraphics[width=0.30\textwidth]{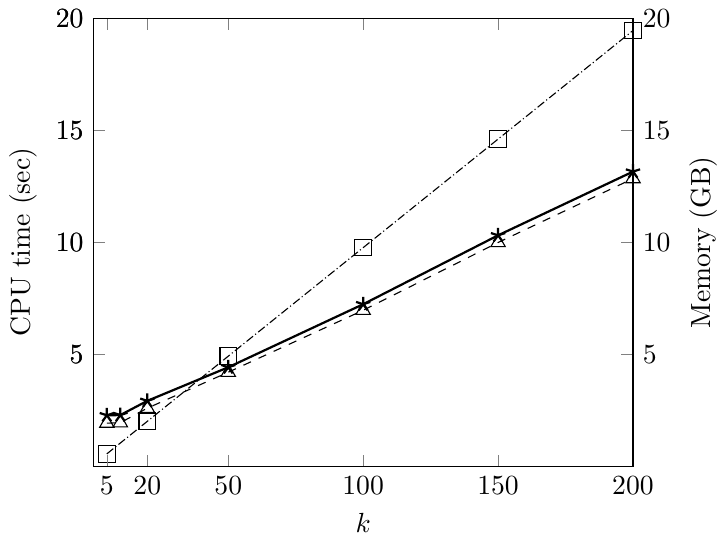}
    }\!\!\!\!
  \subfigure[CTU Network]{
   \label{fig:exp:ip}
    \includegraphics[width=0.31\textwidth]{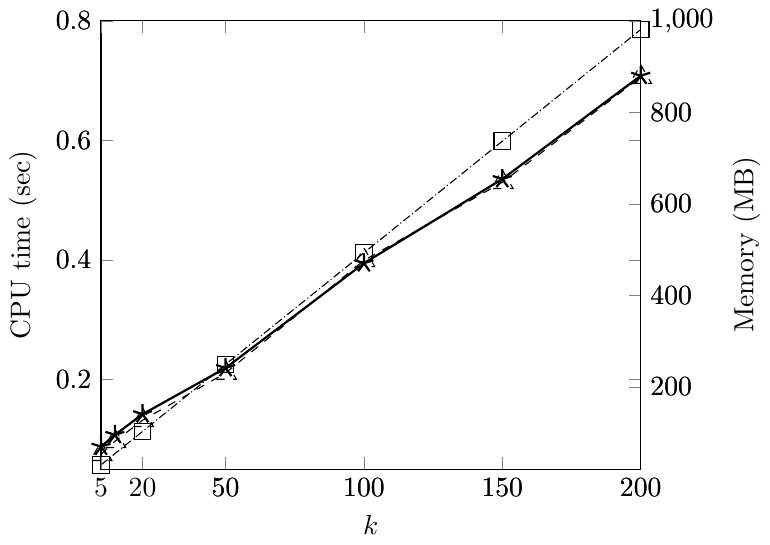}
    }\!\!\!\!
  \subfigure[Prosper Loans Network]{
   \label{fig:exp:prosper}
    \includegraphics[width=0.31\textwidth]{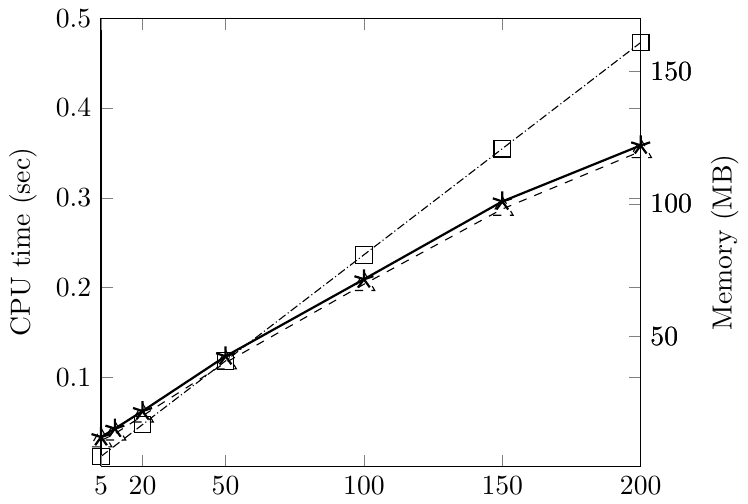}
    }
  \vspace{-0.1in}
  \caption{Selective and grouped proportional provenance}
  \label{fig:exp:sel}
\end{figure*}

Figure \ref{fig:exp:sel} shows the runtime performance (in sec.) and memory
requirements (in MB) for the different values of $k$ on the different
datasets.
As expected the runtime and the memory requirements are roughly
proportional to $k$. For small values of $k$ (less than 20) the
runtime is roughly constant with respect to $k$ (see Figure
\ref{fig:exp:sel}(a)).
This is because of the effect
of SIMD instructions, which make vector operations (lines 9 and 10 of
Algorithm \ref{algo:proportional}) unaffected by the vector size.
SIMD data parallelism is already in full action for values of $k$
greater than $20$, so we observe linear scalability from thereon.

\subsection{Limiting the scope of provenance tracking }\label{sec:exp:scope}
As shown in Section \ref{sec:exp:perf}, proportional provenance tracking 
throughout the entire history of interactions is infeasible, due to
its high memory requirements. In addition, keeping and updating sparse
representations of provenance vectors
becomes expensive over time as the lists grow larger because of the
higher cost of merging operations.

\begin{figure*}[t!]
\includegraphics[width=0.30\textwidth]{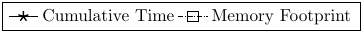}  \\
\subfigure[Bitcoin Network]{
   \label{fig:exp:cum:bitcoin}
    \includegraphics[width=0.30\textwidth]{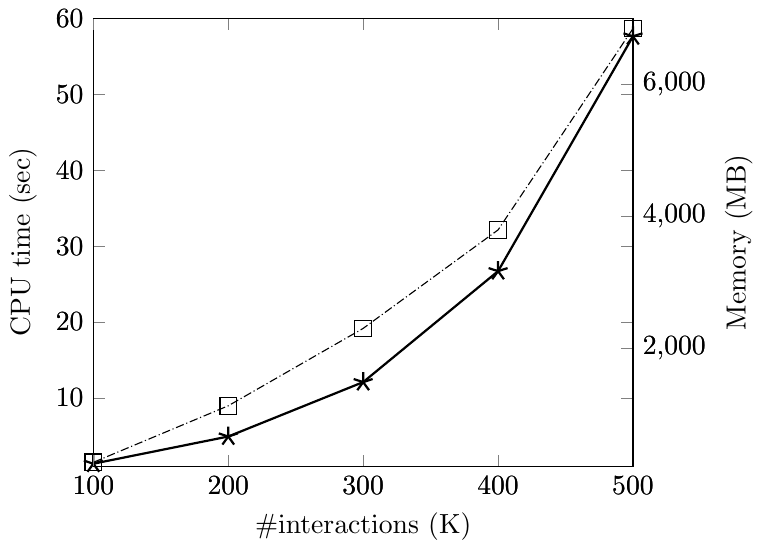}
    }\!\!\!\!
  \subfigure[CTU Network]{
   \label{fig:exp:cum:ip}
    \includegraphics[width=0.30\textwidth]{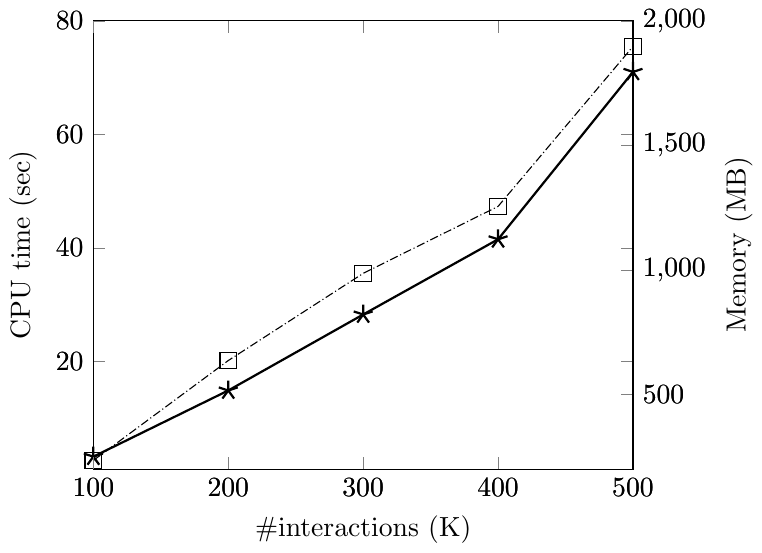}
    }\!\!\!\!
  \subfigure[Prosper Loans Network]{
   \label{fig:exp:cum:prosper}
    \includegraphics[width=0.30\textwidth]{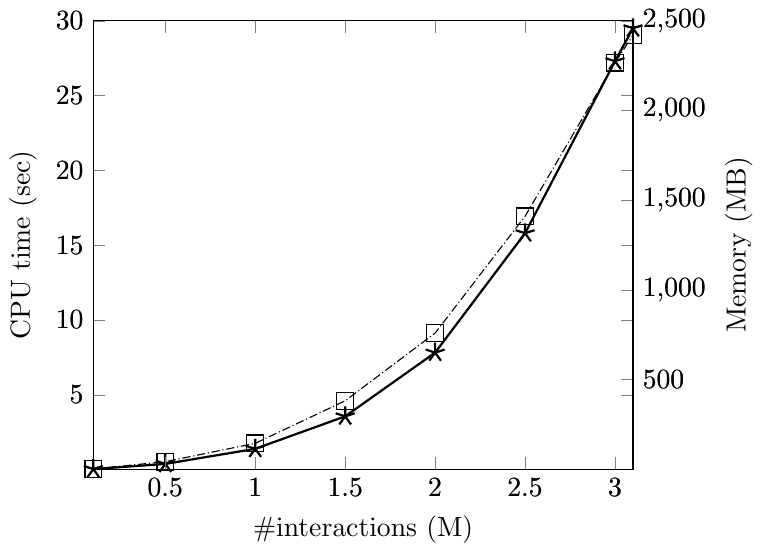}
    }
  \vspace{-0.1in}
  \caption{Cumulative time vs. number of
    processed interactions}
  \label{fig:exp:cum}
\end{figure*}

%The experiments in Figure  \ref{fig:exp:cum} 
%%and
%%\ref{fig:exp:avgtime}
%verify this assertion.
Figure \ref{fig:exp:cum} 
verifies this assertion, by
showing the cumulative time and memory requirements while tracking
proportional selection after each interaction for the first 500K
interactions in Bitcoin and CTU (after this point, the memory
requirements become too high), and for all interactions in Prosper Loans.
Observe that the cumulative runtime
increases superlinearly with the number of interactions and so do the
memory requirements (these two are correlated). 
%Figure
%\ref{fig:exp:avgtime} shows the average runtime cost for each
%interaction in the first 100K interactions, in the second 100K
%interactions, and so on until the first 500K interactions for  Bitcoin
%and CTU and until the last interaction in Prosper Loans. 
%Observe
%that 
The average cost for handling each interaction grows as the
number of processed interactions increases, which is attributed to the
population of the sparse lists that keep the provenance information
for each vertex; merging operations on these lists become
expensive as they grow.

%\begin{figure*}[t!]
%\subfigure[Bitcoin Network]{
%   \label{fig:exp:avg:bitcoin}
%    \includegraphics[width=0.30\textwidth]{plots/interactions_per_time_bc.pdf}
%    }\!\!\!\!
%  \subfigure[CTU Network]{
%   \label{fig:exp:avg:ip}
%    \includegraphics[width=0.30\textwidth]{plots/interactions_per_time_ip.pdf}
%    }\!\!\!\!
%  \subfigure[Prosper Loans Network]{
%   \label{fig:exp:avg:prosper}
%    \includegraphics[width=0.30\textwidth]{plots/interactions_per_time_prosper.pdf}
%    }
%  \vspace{-0.1in}
%  \caption{Average time per interaction vs. number of
%    processed interactions}
%  \label{fig:exp:avgtime}
%\end{figure*}

\begin{figure*}[t!]
\includegraphics[width=0.25\textwidth]{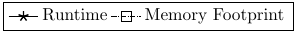}\\    
\subfigure[Bitcoin Network]{
   \label{fig:exp:win:bitcoin}
    \includegraphics[width=0.30\textwidth]{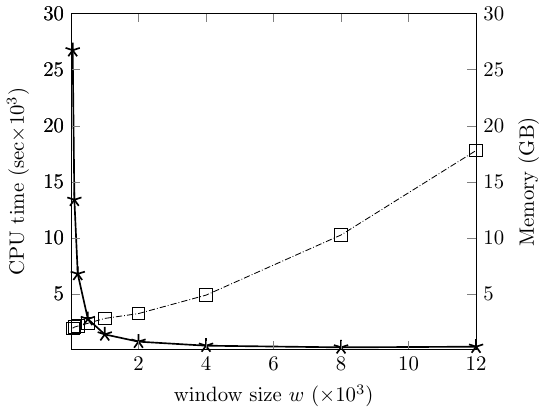}
    }\!\!
  \subfigure[CTU Network]{
   \label{fig:exp:win:ip}
    \includegraphics[width=0.31\textwidth]{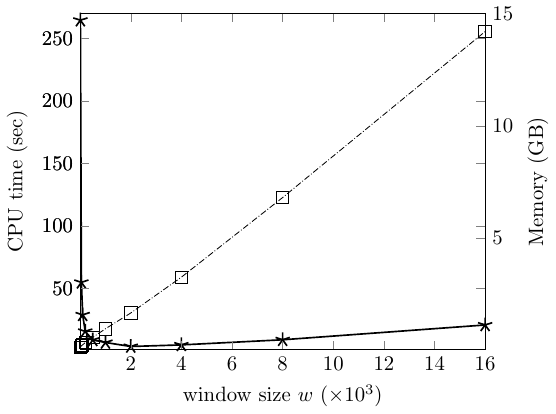}
    }\!\!
  \subfigure[Prosper Loans Network]{
   \label{fig:exp:win:prosper}
    \includegraphics[width=0.30\textwidth]{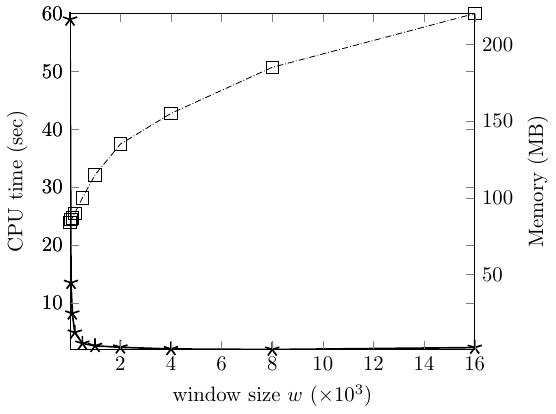}
    }
  \vspace{-0.1in}
  \caption{Windowing approach}
  \label{fig:exp:window}
\end{figure*}

\begin{figure*}[t!]
\includegraphics[width=0.25\textwidth]{plots/legend3.pdf} \\     
\subfigure[Bitcoin Network]{
   \label{fig:exp:bud:bitcoin}
    \includegraphics[width=0.30\textwidth]{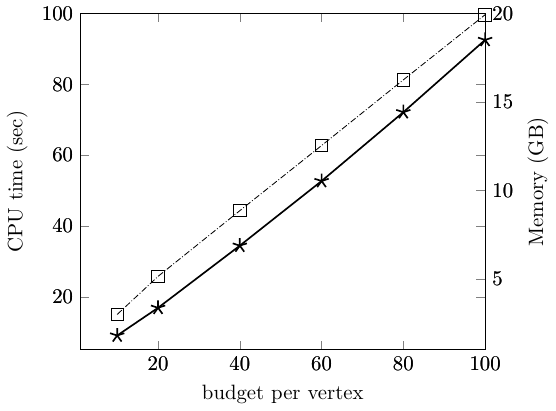}
    }\!\!
  \subfigure[CTU Network]{
   \label{fig:exp:bud:ip}
    \includegraphics[width=0.31\textwidth]{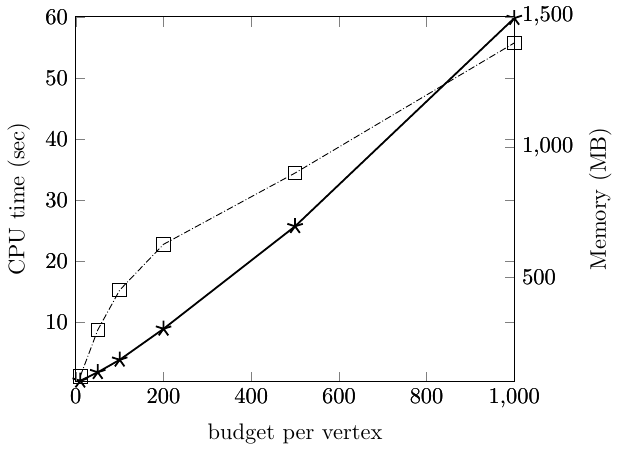}
    }\!\!
  \subfigure[Prosper Loans Network]{
   \label{fig:exp:bud:prosper}
    \includegraphics[width=0.30\textwidth]{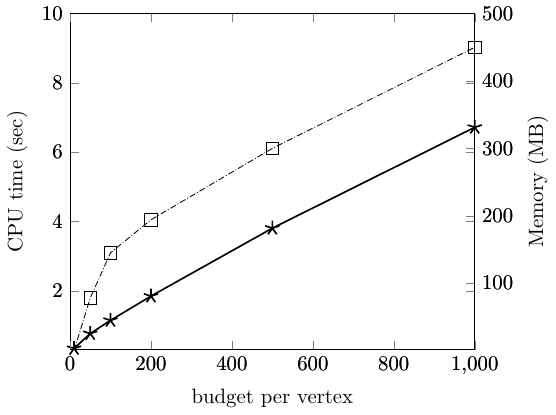}
    }
  \vspace{-0.1in}
  \caption{Budget-based provenance}
  \label{fig:exp:budget}
\end{figure*}

We now evaluate the solutions proposed 
in Section \ref{sec:streaming} for limiting the scope of provenance
tracking in order to make the maintenance of proportional provenance
vectors feasible for large graphs.
Once again, we experimented with the three largest networks and
applied the two approaches proposed in Section \ref{sec:streaming} on
them. 
Figure \ref{fig:exp:window} shows the runtime cost and the memory
requirements of the windowing approach for different values of the
window parameter $W$. As the figure shows, by increasing the size of
the window, we improve the runtime performance as the buffers have to
be reset less frequently. On the other hand, increasing the window
size naturally increases the memory requirements.
For Bitcoin  and Prosper Loans, larger window sizes are
affordable, as the memory requirements do not increase a lot.
On the other hand, for CTU the memory requirements almost double
when $W$ doubles.
In summary, the windowing technique is very useful, especially when we
want to have guaranteed accurate provenance information for all
vertices up to a
time window in the past.
%We should use as large windows as the system's memory permits,
%as this reduces the number of times when we need to perform buffer resets.
%\todo{we need experiments with larger window sizes, these can run on
 % the UOI machines, as memories are sufficiently big}

Figure \ref{fig:exp:budget} shows the runtime cost and the memory
requirements of the budget-based approach for different values of the
maximum budget $C$ given as capacity for provenance entries to each
vertex.
As the figure shows, by increasing the budget $C$ per vertex, the
runtime cost to maintain provenance increases, as the provenance
information at buffers becomes larger and merging lists becomes more
expensive. The increase in the runtime cost is not very high though,
because many lists remain relatively short and the number of list
shrinks are less frequent. 
At the same time, the space requirements grow linearly with $C$, which
means that very large values of $C$ are not affordable for large
graphs like Bitcoin.

In order to assess the value of this approach, in Table \ref{tab:exp:shrink}, 
%we also measured the
%average number of times each non-empty buffer is shrunk throughout the timeline
%as a function of $C$.
%Table \ref{tab:exp:shrink} shows some statistics with respect to this.
we measured  for each of the three large datasets and for different values%
\footnote{We could not use values of $C$ larger than 100 on Bitcoin due to
  memory constraints.}
 of $C$,
(i) the number of times each non-empty buffer has been
shrunk and (ii) the percentage of vertices (with non-empty buffer)
whose buffer was shrunk at least once.
Especially for the larger networks with high memory requirements
(Bitcoin and CTU), we observe that the number of shrinks and the
percentage of vertices where they take place converge to low values
and, after some point, increasing $C$ does not offer much benefit.
Overall, the budget-based approach is attractive since each buffer is
shrunk only a few times on average, meaning that the provenance
information loss is limited even in large graphs.
For example, at the Bitcoin network, for a value of $C=50$, each
buffer is shrunk 1.5 times on average after 45M interactions, meaning
that each buffer tracks provenance information that traces back to
tens of millions of transactions before.

\begin{table}[ht]
  \scriptsize
  \caption{Shrinking statistics in budget-based provenance}
  \label{tab:exp:shrink}
  \centering
%  \scriptsize
  \vspace{-0.2cm}
  \begin{tabular}{|l|c|c|c|c|c|c|}
    \hline
$C$&\multicolumn{2}{|c|}{Bitcoin Network}&\multicolumn{2}{|c|}{CTU Network}&\multicolumn{2}{|c|}{Prosper Loans Network}\\
\hline
&avg. shrinks&\% vertices&avg. shrinks&\% vertices&avg. shrinks&\% vertices\\
\hline
 10&1.94&18.38&7.27&31.07&20.67&94.7\\
50&1.51&14.79&5.1&28.68&4.77&79.29\\
100&1.43&14.21&4.77&27.94&2.97&69.09\\
200&--&--&4.53&26.6&2.1&59.16\\
500&--&--&4.34&25.24&1.5&47.64\\
                      1000&--&--&4.3&25.02&1.23&41.39                                            \\
\hline
\end{tabular}
\end{table}

\subsection{Path tracking }\label{sec:exp:paths}
As discussed in Section \ref{sec:howprovenance}, in might be
desirable, for each quantity that has reached the buffer of a vertex
to know not just the vertex that generated the quantity, but also the
path that the quantity followed in the graph until it reached its
destination.
In the next experiment, we evaluate the overhead of tracking the paths
(i.e., {\em how}-provenance) compared to just tracking the origins of
the quantities. We implemented path tracking as part of the LIFO
selection policy for provenance (Section
\ref{sec:models:receipt})
and used it to track the paths for all (origin, quantity) pairs accumulated at
vertices after processing all interactions in all datasets.
Table \ref{tab:exp:paths} shows the runtime performance, the
memory requirements, and the average path length for each quantity
element. The memory requirements are split
into the memory required to store the provenance entries in the lists
(as in LIFO) and the memory required to store the paths.
Observe that for most datasets the memory overhead for keeping the
paths is not extremely high.
This overhead is determined by the average path
length (last column of the table), which is relatively low in four out
of the five datasets.
Only in Flights the storage overhead for the
paths is very high because quantities travel a long way.  
In this dataset, the number of vertices is very small compared to the
number of interactions, so we can expect very long paths.
Still, on all datasets, the runtime is only up to a few times higher
compared to tracking just the origins and not the paths (see Table
\ref{table:runtimeall}, column LIFO), meaning that path tracking is
feasible even for very long sequences of interactions on large graphs,
like Bitcoin.

\begin{table}[ht]
  \scriptsize
  \caption{Tracking provenance paths in LIFO}
  \label{tab:exp:paths}
  \centering
%  \scriptsize
  \vspace{-0.2cm}
  \begin{tabular}{|l|@{~}c@{~}|@{~}c@{~}|@{~}c@{~}|@{~}c@{~}|@{~}c@{~}|}
    \hline
 Dataset   &time (sec.)&mem entries (MB)&mem paths (MB) &total mem  (MB)  &avg. path length\\
    \hline
Bitcoin&13.35&534.62&847.50&1382.13&4.75\\
CTU&0.36&33.87&7.16&41.03&0.63\\
Prosper&0.4&36.85&0.74&37.59&0.06\\
Flights&0.17&0.627&57.09&57.72&273.17\\
Taxis&0.008&0.58&1.09&1.68&5.55\\
\hline
\end{tabular}
\end{table}

\vspace{-5mm}
\subsection{Use case}\label{sec:exp:usecase}
\begin{figure}[h]
	\centering
	\includegraphics[width=0.70\columnwidth]{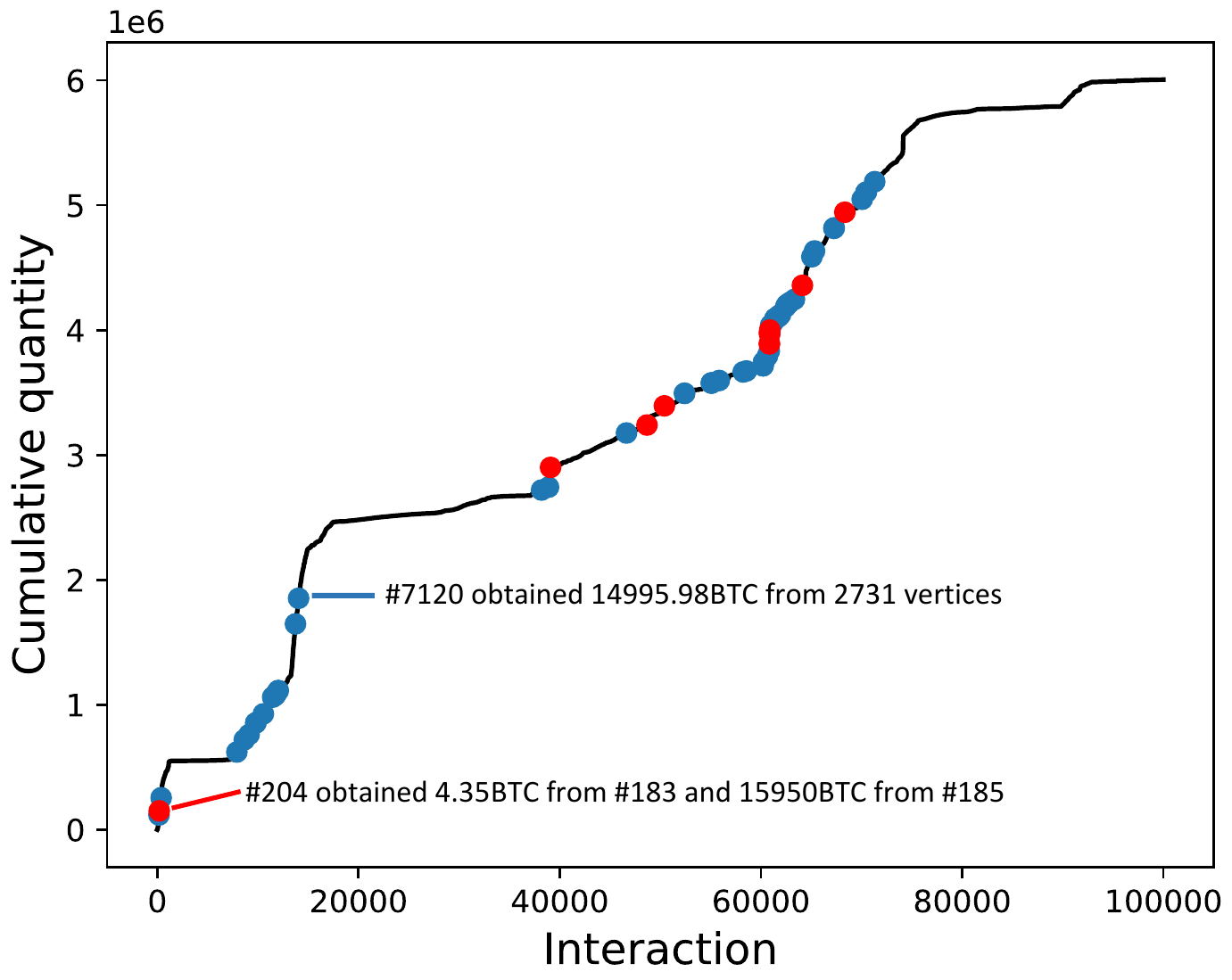}
	\caption{Provenance alerts in Bitcoin}
	\label{fig:alert}
\end{figure}

Figure \ref{fig:alert} demonstrates a practical example of provenance tracing in TINs.
The plot shows the total accumulated quantities at the vertices of Bitcoin after each interaction (first 100K interactions, proportional selection policy).
Consider a data analyst who wants to be alerted whenever a vertex $v$ accumulates a significant amount of money, which does not originate from $v$'s direct neighbors (i.e., $v$'s neighbors just relay amounts to $v$). Hence, after each interaction, we issue an alert when the receiving vertex does not have any quantity that originates from its neighbors and the total quantity in its buffer exceeds 10K BTC. The colored dots in the figure show these alerts (89 in total) and provenance information for some of them. Red dots are alerts where the number of contributing vertices is less than five (the rest of them are blue). We observe that in most cases the amount was received from numerous vertices (an indication of possible ``smurfing'').
This alerting mechanism is very efficient and easy to implement, as we only have to maintain at each vertex $v$ the total quantity that originates from vertices that transfer quantities to $v$ (i.e., direct neighbors of $v$).

\section{Conclusions}\label{sec:conclusion}
In this paper, we introduced and studied provenance in temporal
interaction networks (TINs). To the best of our knowledge, we are the first
to define and study this problem, considering the data transfers
among the vertices, as interactions take place over time.
We investigate different selection policies for data propagation
in TINs that correspond to different application scenaria.
For each policy, we propose propagation mechanisms for provenance
(annotation) data and analyze their space and time complexities.
For the hardest policy (proportional selection), we
propose to track provenance from a limited set of vertices or
from groups thereof. We also propose to
limit provenance tracking up to a sliding window of past interactions
or to set a space budget at each vertex for provenance tracking.
We evaluated our methods using
five real datasets and demonstrated their scalability.
In the future, we plan to investigate {\em lazy} approaches for data
provenance in TINs (e.g., apply the replay lazy \cite{GlavicEFT13}
approach, or investigate backtracing methods).
In addition, we plan to study whether our approaches for TINs can be
adapted to be applied on social networks, where data are
diffused, instead on being relayed from vertex to vertex.
Finally, we plan to analyze in depth the computed provenance data in
TINs, with the help of data mining approaches, in order find  interesting insights in them.

\newpage
\balance

\bibliographystyle{ACM-Reference-Format}
\bibliography{references}


\begin{thebibliography}{47}

%%% the Web address.

\ifx \showCODEN    \undefined \def \showCODEN     #1{\unskip}     \fi
\ifx \showDOI      \undefined \def \showDOI       #1{#1}\fi
\ifx \showISBNx    \undefined \def \showISBNx     #1{\unskip}     \fi
\ifx \showISBNxiii \undefined \def \showISBNxiii  #1{\unskip}     \fi
\ifx \showISSN     \undefined \def \showISSN      #1{\unskip}     \fi
\ifx \showLCCN     \undefined \def \showLCCN      #1{\unskip}     \fi
\ifx \shownote     \undefined \def \shownote      #1{#1}          \fi
\ifx \showarticletitle \undefined \def \showarticletitle #1{#1}   \fi
\ifx \showURL      \undefined \def \showURL       {\relax}        \fi
% invisible to TeX
\providecommand\bibfield[2]{#2}
\providecommand\bibinfo[2]{#2}
\providecommand\natexlab[1]{#1}
\providecommand\showeprint[2][]{arXiv:#2}

\bibitem[\protect\citeauthoryear{Agrawal, Benjelloun, Sarma, Hayworth, Nabar,
  Sugihara, and Widom}{Agrawal et~al\mbox{.}}{2006}]%
        {DBLP:conf/vldb/AgrawalBSHNSW06}
\bibfield{author}{\bibinfo{person}{Parag Agrawal}, \bibinfo{person}{Omar
  Benjelloun}, \bibinfo{person}{Anish~Das Sarma}, \bibinfo{person}{Chris
  Hayworth}, \bibinfo{person}{Shubha~U. Nabar}, \bibinfo{person}{Tomoe
  Sugihara}, {and} \bibinfo{person}{Jennifer Widom}.}
  \bibinfo{year}{2006}\natexlab{}.
\newblock \showarticletitle{Trio: {A} System for Data, Uncertainty, and
  Lineage}. In \bibinfo{booktitle}{\emph{Proceedings of the 32nd International
  Conference on Very Large Data Bases, Seoul, Korea, September 12-15, 2006}}.
  \bibinfo{pages}{1151--1154}.
\newblock


\bibitem[\protect\citeauthoryear{Akrida, Czyzowicz, Gasieniec, Kuszner, and
  Spirakis}{Akrida et~al\mbox{.}}{2017}]%
        {DBLP:conf/ciac/AkridaCGKS17}
\bibfield{author}{\bibinfo{person}{Eleni~C. Akrida}, \bibinfo{person}{Jurek
  Czyzowicz}, \bibinfo{person}{Leszek Gasieniec}, \bibinfo{person}{Lukasz
  Kuszner}, {and} \bibinfo{person}{Paul~G. Spirakis}.}
  \bibinfo{year}{2017}\natexlab{}.
\newblock \showarticletitle{Temporal Flows in Temporal Networks}. In
  \bibinfo{booktitle}{\emph{Algorithms and Complexity - 10th International
  Conference, {CIAC} 2017, Athens, Greece, May 24-26, 2017, Proceedings}}.
  \bibinfo{pages}{43--54}.
\newblock


\bibitem[\protect\citeauthoryear{Barbier, Feng, Gundecha, and Liu}{Barbier
  et~al\mbox{.}}{2013}]%
        {barbier2013provenance}
\bibfield{author}{\bibinfo{person}{Geoffrey Barbier}, \bibinfo{person}{Zhuo
  Feng}, \bibinfo{person}{Pritam Gundecha}, {and} \bibinfo{person}{Huan Liu}.}
  \bibinfo{year}{2013}\natexlab{}.
\newblock \showarticletitle{Provenance data in social media}.
\newblock \bibinfo{journal}{\emph{Synthesis Lectures on Data Mining and
  Knowledge Discovery}} (\bibinfo{year}{2013}), \bibinfo{pages}{1--84}.
\newblock


\bibitem[\protect\citeauthoryear{Belth, Zheng, and Koutra}{Belth
  et~al\mbox{.}}{[n.d.]}]%
        {DBLP:conf/kdd/BelthZK20}
\bibfield{author}{\bibinfo{person}{Caleb Belth}, \bibinfo{person}{Xinyi Zheng},
  {and} \bibinfo{person}{Danai Koutra}.} \bibinfo{year}{[n.d.]}\natexlab{}.
\newblock \showarticletitle{Mining Persistent Activity in Continually Evolving
  Networks}. In \bibinfo{booktitle}{\emph{{KDD} '20: The 26th {ACM} {SIGKDD}
  Conference on Knowledge Discovery and Data Mining, Virtual Event, CA, USA,
  August 23-27, 2020}}, \bibfield{editor}{\bibinfo{person}{Rajesh Gupta},
  \bibinfo{person}{Yan Liu}, \bibinfo{person}{Jiliang Tang}, {and}
  \bibinfo{person}{B.~Aditya Prakash}} (Eds.). \bibinfo{pages}{934--944}.
\newblock


\bibitem[\protect\citeauthoryear{Bhagwat, Chiticariu, Tan, and
  Vijayvargiya}{Bhagwat et~al\mbox{.}}{2004}]%
        {DBLP:conf/vldb/BhagwatCTV04}
\bibfield{author}{\bibinfo{person}{Deepavali Bhagwat}, \bibinfo{person}{Laura
  Chiticariu}, \bibinfo{person}{Wang~Chiew Tan}, {and} \bibinfo{person}{Gaurav
  Vijayvargiya}.} \bibinfo{year}{2004}\natexlab{}.
\newblock \showarticletitle{An Annotation Management System for Relational
  Databases}. In \bibinfo{booktitle}{\emph{(e)Proceedings of the Thirtieth
  International Conference on Very Large Data Bases, {VLDB} 2004, Toronto,
  Canada, August 31 - September 3 2004}}. \bibinfo{pages}{900--911}.
\newblock


\bibitem[\protect\citeauthoryear{Buneman, Khanna, and Tan}{Buneman
  et~al\mbox{.}}{2001}]%
        {DBLP:conf/icdt/BunemanKT01}
\bibfield{author}{\bibinfo{person}{Peter Buneman}, \bibinfo{person}{Sanjeev
  Khanna}, {and} \bibinfo{person}{Wang~Chiew Tan}.}
  \bibinfo{year}{2001}\natexlab{}.
\newblock \showarticletitle{Why and Where: {A} Characterization of Data
  Provenance}. In \bibinfo{booktitle}{\emph{Database Theory - {ICDT} 2001, 8th
  International Conference}}. \bibinfo{pages}{316--330}.
\newblock


\bibitem[\protect\citeauthoryear{Buneman, Khanna, and Tan}{Buneman
  et~al\mbox{.}}{2002}]%
        {BunemanKT02}
\bibfield{author}{\bibinfo{person}{Peter Buneman}, \bibinfo{person}{Sanjeev
  Khanna}, {and} \bibinfo{person}{Wang~Chiew Tan}.}
  \bibinfo{year}{2002}\natexlab{}.
\newblock \showarticletitle{On Propagation of Deletions and Annotations Through
  Views}. In \bibinfo{booktitle}{\emph{{PODS}}}. \bibinfo{pages}{150--158}.
\newblock


\bibitem[\protect\citeauthoryear{Chapman, Jagadish, and Ramanan}{Chapman
  et~al\mbox{.}}{2008}]%
        {DBLP:conf/sigmod/ChapmanJR08}
\bibfield{author}{\bibinfo{person}{Adriane Chapman}, \bibinfo{person}{H.~V.
  Jagadish}, {and} \bibinfo{person}{Prakash Ramanan}.}
  \bibinfo{year}{2008}\natexlab{}.
\newblock \showarticletitle{Efficient provenance storage}. In
  \bibinfo{booktitle}{\emph{Proceedings of the {ACM} {SIGMOD} International
  Conference on Management of Data, {SIGMOD} 2008, Vancouver, BC, Canada, June
  10-12, 2008}}. \bibinfo{pages}{993--1006}.
\newblock


\bibitem[\protect\citeauthoryear{Cheney, Chiticariu, and Tan}{Cheney
  et~al\mbox{.}}{2009}]%
        {DBLP:journals/ftdb/CheneyCT09}
\bibfield{author}{\bibinfo{person}{James Cheney}, \bibinfo{person}{Laura
  Chiticariu}, {and} \bibinfo{person}{Wang~Chiew Tan}.}
  \bibinfo{year}{2009}\natexlab{}.
\newblock \showarticletitle{Provenance in Databases: Why, How, and Where}.
\newblock \bibinfo{journal}{\emph{Found. Trends Databases}}
  (\bibinfo{year}{2009}), \bibinfo{pages}{379--474}.
\newblock


\bibitem[\protect\citeauthoryear{Chothia, Liagouris, McSherry, and
  Roscoe}{Chothia et~al\mbox{.}}{2016}]%
        {DBLP:journals/pvldb/ChothiaLMR16}
\bibfield{author}{\bibinfo{person}{Zaheer Chothia}, \bibinfo{person}{John
  Liagouris}, \bibinfo{person}{Frank McSherry}, {and} \bibinfo{person}{Timothy
  Roscoe}.} \bibinfo{year}{2016}\natexlab{}.
\newblock \showarticletitle{Explaining Outputs in Modern Data Analytics}.
\newblock \bibinfo{journal}{\emph{Proc. {VLDB} Endow.}} (\bibinfo{year}{2016}),
  \bibinfo{pages}{1137--1148}.
\newblock


\bibitem[\protect\citeauthoryear{de~Paula, Holanda, Gomes, Lifschitz, and
  Walter}{de~Paula et~al\mbox{.}}{2013}]%
        {DBLP:journals/bmcbi/PaulaHGLW13}
\bibfield{author}{\bibinfo{person}{Renato de Paula}, \bibinfo{person}{Maristela
  Holanda}, \bibinfo{person}{Luciana S.~A. Gomes},
  \bibinfo{person}{S{\'{e}}rgio Lifschitz}, {and} \bibinfo{person}{Maria
  Em{\'{\i}}lia M.~T. Walter}.} \bibinfo{year}{2013}\natexlab{}.
\newblock \showarticletitle{Provenance in bioinformatics workflows}.
\newblock \bibinfo{journal}{\emph{{BMC} Bioinform.}} (\bibinfo{year}{2013}),
  \bibinfo{pages}{S6}.
\newblock


\bibitem[\protect\citeauthoryear{Decker and Wattenhofer}{Decker and
  Wattenhofer}{2013}]%
        {DBLP:conf/p2p/DeckerW13}
\bibfield{author}{\bibinfo{person}{Christian Decker} {and}
  \bibinfo{person}{Roger Wattenhofer}.} \bibinfo{year}{2013}\natexlab{}.
\newblock \showarticletitle{Information propagation in the Bitcoin network}. In
  \bibinfo{booktitle}{\emph{13th {IEEE} International Conference on
  Peer-to-Peer Computing}}. \bibinfo{pages}{1--10}.
\newblock


\bibitem[\protect\citeauthoryear{Domingos and Richardson}{Domingos and
  Richardson}{2001}]%
        {DBLP:conf/kdd/DomingosR01}
\bibfield{author}{\bibinfo{person}{Pedro~M. Domingos} {and}
  \bibinfo{person}{Matthew Richardson}.} \bibinfo{year}{2001}\natexlab{}.
\newblock \showarticletitle{Mining the network value of customers}. In
  \bibinfo{booktitle}{\emph{Proceedings of the seventh {ACM} {SIGKDD}
  international conference on Knowledge discovery and data mining}}.
  \bibinfo{pages}{57--66}.
\newblock


\bibitem[\protect\citeauthoryear{Garc{\'{\i}}a, Grill, Stiborek, and
  Zunino}{Garc{\'{\i}}a et~al\mbox{.}}{2014}]%
        {DBLP:journals/compsec/GarciaGSZ14}
\bibfield{author}{\bibinfo{person}{Sebasti{\'{a}}n Garc{\'{\i}}a},
  \bibinfo{person}{Martin Grill}, \bibinfo{person}{Jan Stiborek}, {and}
  \bibinfo{person}{Alejandro Zunino}.} \bibinfo{year}{2014}\natexlab{}.
\newblock \showarticletitle{An empirical comparison of botnet detection
  methods}.
\newblock \bibinfo{journal}{\emph{Comput. Secur.}} (\bibinfo{year}{2014}),
  \bibinfo{pages}{100--123}.
\newblock


\bibitem[\protect\citeauthoryear{Geerts, Kementsietsidis, and Milano}{Geerts
  et~al\mbox{.}}{2006}]%
        {DBLP:conf/icde/GeertsKM06}
\bibfield{author}{\bibinfo{person}{Floris Geerts}, \bibinfo{person}{Anastasios
  Kementsietsidis}, {and} \bibinfo{person}{Diego Milano}.}
  \bibinfo{year}{2006}\natexlab{}.
\newblock \showarticletitle{{MONDRIAN:} Annotating and Querying Databases
  through Colors and Blocks}. In \bibinfo{booktitle}{\emph{Proceedings of the
  22nd International Conference on Data Engineering, {ICDE} 2006, 3-8 April
  2006, Atlanta, GA, {USA}}}. \bibinfo{pages}{82}.
\newblock


\bibitem[\protect\citeauthoryear{Glavic, Esmaili, Fischer, and Tatbul}{Glavic
  et~al\mbox{.}}{2013}]%
        {GlavicEFT13}
\bibfield{author}{\bibinfo{person}{Boris Glavic},
  \bibinfo{person}{Kyumars~Sheykh Esmaili}, \bibinfo{person}{Peter~Michael
  Fischer}, {and} \bibinfo{person}{Nesime Tatbul}.}
  \bibinfo{year}{2013}\natexlab{}.
\newblock \showarticletitle{Ariadne: managing fine-grained provenance on data
  streams}. In \bibinfo{booktitle}{\emph{{DEBS}}}. \bibinfo{publisher}{{ACM}},
  \bibinfo{pages}{39--50}.
\newblock


\bibitem[\protect\citeauthoryear{Green, Karvounarakis, Ives, and Tannen}{Green
  et~al\mbox{.}}{2010}]%
        {DBLP:journals/debu/GreenKIT10}
\bibfield{author}{\bibinfo{person}{Todd~J. Green}, \bibinfo{person}{Grigoris
  Karvounarakis}, \bibinfo{person}{Zachary~G. Ives}, {and} \bibinfo{person}{Val
  Tannen}.} \bibinfo{year}{2010}\natexlab{}.
\newblock \showarticletitle{Provenance in {ORCHESTRA}}.
\newblock \bibinfo{journal}{\emph{{IEEE} Data Eng. Bull.}}
  (\bibinfo{year}{2010}), \bibinfo{pages}{9--16}.
\newblock


\bibitem[\protect\citeauthoryear{Gundecha, Feng, and Liu}{Gundecha
  et~al\mbox{.}}{2013}]%
        {DBLP:conf/cikm/GundechaFL13}
\bibfield{author}{\bibinfo{person}{Pritam Gundecha}, \bibinfo{person}{Zhuo
  Feng}, {and} \bibinfo{person}{Huan Liu}.} \bibinfo{year}{2013}\natexlab{}.
\newblock \showarticletitle{Seeking provenance of information using social
  media}. In \bibinfo{booktitle}{\emph{22nd {ACM} International Conference on
  Information and Knowledge Management, CIKM}}. \bibinfo{pages}{1691--1696}.
\newblock


\bibitem[\protect\citeauthoryear{Han, Pasquier, Bates, Mickens, and
  Seltzer}{Han et~al\mbox{.}}{2020}]%
        {HanP0MS20}
\bibfield{author}{\bibinfo{person}{Xueyuan Han}, \bibinfo{person}{Thomas
  F.~J.{-}M. Pasquier}, \bibinfo{person}{Adam Bates}, \bibinfo{person}{James
  Mickens}, {and} \bibinfo{person}{Margo~I. Seltzer}.}
  \bibinfo{year}{2020}\natexlab{}.
\newblock \showarticletitle{Unicorn: Runtime Provenance-Based Detector for
  Advanced Persistent Threats}. In \bibinfo{booktitle}{\emph{27th Annual
  Network and Distributed System Security Symposium, {NDSS} 2020, San Diego,
  California, USA, February 23-26, 2020}}.
\newblock


\bibitem[\protect\citeauthoryear{Heinis and Alonso}{Heinis and Alonso}{2008}]%
        {DBLP:conf/sigmod/HeinisA08}
\bibfield{author}{\bibinfo{person}{Thomas Heinis} {and}
  \bibinfo{person}{Gustavo Alonso}.} \bibinfo{year}{2008}\natexlab{}.
\newblock \showarticletitle{Efficient lineage tracking for scientific
  workflows}. In \bibinfo{booktitle}{\emph{Proceedings of the {ACM} {SIGMOD}
  International Conference on Management of Data, {SIGMOD} 2008, Vancouver, BC,
  Canada, June 10-12, 2008}}. \bibinfo{pages}{1007--1018}.
\newblock


\bibitem[\protect\citeauthoryear{Holme and Saram{\"a}ki}{Holme and
  Saram{\"a}ki}{2012}]%
        {holme2012temporal}
\bibfield{author}{\bibinfo{person}{Petter Holme} {and} \bibinfo{person}{Jari
  Saram{\"a}ki}.} \bibinfo{year}{2012}\natexlab{}.
\newblock \showarticletitle{Temporal networks}.
\newblock \bibinfo{journal}{\emph{Physics reports}} (\bibinfo{year}{2012}),
  \bibinfo{pages}{97--125}.
\newblock


\bibitem[\protect\citeauthoryear{Interlandi, Shah, Tetali, Gulzar, Yoo, Kim,
  Millstein, and Condie}{Interlandi et~al\mbox{.}}{2015}]%
        {DBLP:journals/pvldb/InterlandiSTGYK15}
\bibfield{author}{\bibinfo{person}{Matteo Interlandi}, \bibinfo{person}{Kshitij
  Shah}, \bibinfo{person}{Sai~Deep Tetali}, \bibinfo{person}{Muhammad~Ali
  Gulzar}, \bibinfo{person}{Seunghyun Yoo}, \bibinfo{person}{Miryung Kim},
  \bibinfo{person}{Todd~D. Millstein}, {and} \bibinfo{person}{Tyson Condie}.}
  \bibinfo{year}{2015}\natexlab{}.
\newblock \showarticletitle{Titian: Data Provenance Support in Spark}.
\newblock \bibinfo{journal}{\emph{Proc. {VLDB} Endow.}} (\bibinfo{year}{2015}),
  \bibinfo{pages}{216--227}.
\newblock


\bibitem[\protect\citeauthoryear{Karvounarakis, Ives, and Tannen}{Karvounarakis
  et~al\mbox{.}}{2010}]%
        {DBLP:conf/sigmod/KarvounarakisIT10}
\bibfield{author}{\bibinfo{person}{Grigoris Karvounarakis},
  \bibinfo{person}{Zachary~G. Ives}, {and} \bibinfo{person}{Val Tannen}.}
  \bibinfo{year}{2010}\natexlab{}.
\newblock \showarticletitle{Querying data provenance}. In
  \bibinfo{booktitle}{\emph{Proceedings of the {ACM} {SIGMOD} International
  Conference on Management of Data, {SIGMOD} 2010, Indianapolis, Indiana, USA,
  June 6-10, 2010}}. \bibinfo{pages}{951--962}.
\newblock


\bibitem[\protect\citeauthoryear{Karypis and Kumar}{Karypis and Kumar}{1998}]%
        {DBLP:journals/siamsc/KarypisK98}
\bibfield{author}{\bibinfo{person}{George Karypis} {and} \bibinfo{person}{Vipin
  Kumar}.} \bibinfo{year}{1998}\natexlab{}.
\newblock \showarticletitle{A Fast and High Quality Multilevel Scheme for
  Partitioning Irregular Graphs}.
\newblock \bibinfo{journal}{\emph{{SIAM} J. Sci. Comput.}}
  \bibinfo{volume}{20}, \bibinfo{number}{1} (\bibinfo{year}{1998}),
  \bibinfo{pages}{359--392}.
\newblock


\bibitem[\protect\citeauthoryear{Kempe, Kleinberg, and Tardos}{Kempe
  et~al\mbox{.}}{2003}]%
        {DBLP:conf/kdd/KempeKT03}
\bibfield{author}{\bibinfo{person}{David Kempe}, \bibinfo{person}{Jon~M.
  Kleinberg}, {and} \bibinfo{person}{{\'{E}}va Tardos}.}
  \bibinfo{year}{2003}\natexlab{}.
\newblock \showarticletitle{Maximizing the spread of influence through a social
  network}. In \bibinfo{booktitle}{\emph{Proceedings of the Ninth {ACM}
  {SIGKDD} International Conference on Knowledge Discovery and Data Mining}}.
  \bibinfo{pages}{137--146}.
\newblock


\bibitem[\protect\citeauthoryear{Kondor, P{\'{o}}sfai, Csabai, and
  Vattay}{Kondor et~al\mbox{.}}{2013}]%
        {DBLP:journals/corr/KondorPCV13}
\bibfield{author}{\bibinfo{person}{D{\'{a}}niel Kondor},
  \bibinfo{person}{M{\'{a}}rton P{\'{o}}sfai}, \bibinfo{person}{Istv{\'{a}}n
  Csabai}, {and} \bibinfo{person}{G{\'{a}}bor Vattay}.}
  \bibinfo{year}{2013}\natexlab{}.
\newblock \showarticletitle{Do the rich get richer? An empirical analysis of
  the BitCoin transaction network}.
\newblock \bibinfo{journal}{\emph{CoRR}} (\bibinfo{year}{2013}).
\newblock


\bibitem[\protect\citeauthoryear{Kosyfaki, Mamoulis, Pitoura, and
  Tsaparas}{Kosyfaki et~al\mbox{.}}{2019}]%
        {DBLP:conf/edbt/KosyfakiMPT19}
\bibfield{author}{\bibinfo{person}{Chrysanthi Kosyfaki}, \bibinfo{person}{Nikos
  Mamoulis}, \bibinfo{person}{Evaggelia Pitoura}, {and}
  \bibinfo{person}{Panayiotis Tsaparas}.} \bibinfo{year}{2019}\natexlab{}.
\newblock \showarticletitle{Flow Motifs in Interaction Networks}. In
  \bibinfo{booktitle}{\emph{Advances in Database Technology - 22nd
  International Conference on Extending Database Technology, {EDBT} 2019,
  Lisbon, Portugal, March 26-29, 2019}}. \bibinfo{pages}{241--252}.
\newblock


\bibitem[\protect\citeauthoryear{Kosyfaki, Mamoulis, Pitoura, and
  Tsaparas}{Kosyfaki et~al\mbox{.}}{2021}]%
        {DBLP:conf/icde/KosyfakiMPT21}
\bibfield{author}{\bibinfo{person}{Chrysanthi Kosyfaki}, \bibinfo{person}{Nikos
  Mamoulis}, \bibinfo{person}{Evaggelia Pitoura}, {and}
  \bibinfo{person}{Panayiotis Tsaparas}.} \bibinfo{year}{2021}\natexlab{}.
\newblock \showarticletitle{Flow Computation in Temporal Interaction Networks}.
  In \bibinfo{booktitle}{\emph{37th {IEEE} International Conference on Data
  Engineering, {ICDE} 2021, Chania, Greece, April 19-22, 2021}}.
  \bibinfo{pages}{660--671}.
\newblock


\bibitem[\protect\citeauthoryear{Kumar and Calders}{Kumar and Calders}{2017}]%
        {DBLP:conf/edbt/0002C17}
\bibfield{author}{\bibinfo{person}{Rohit Kumar} {and} \bibinfo{person}{Toon
  Calders}.} \bibinfo{year}{2017}\natexlab{}.
\newblock \showarticletitle{Information Propagation in Interaction Networks}.
  In \bibinfo{booktitle}{\emph{Proceedings of the 20th International Conference
  on Extending Database Technology, {EDBT}}}. \bibinfo{pages}{270--281}.
\newblock


\bibitem[\protect\citeauthoryear{Lee, Lud{\"{a}}scher, and Glavic}{Lee
  et~al\mbox{.}}{2018}]%
        {DBLP:journals/pvldb/LeeLG18}
\bibfield{author}{\bibinfo{person}{Seokki Lee}, \bibinfo{person}{Bertram
  Lud{\"{a}}scher}, {and} \bibinfo{person}{Boris Glavic}.}
  \bibinfo{year}{2018}\natexlab{}.
\newblock \showarticletitle{Provenance Summaries for Answers and Non-Answers}.
\newblock \bibinfo{journal}{\emph{Proc. {VLDB} Endow.}} (\bibinfo{year}{2018}),
  \bibinfo{pages}{1954--1957}.
\newblock


\bibitem[\protect\citeauthoryear{Lee, Lud{\"{a}}scher, and Glavic}{Lee
  et~al\mbox{.}}{2019}]%
        {DBLP:journals/vldb/LeeLG19}
\bibfield{author}{\bibinfo{person}{Seokki Lee}, \bibinfo{person}{Bertram
  Lud{\"{a}}scher}, {and} \bibinfo{person}{Boris Glavic}.}
  \bibinfo{year}{2019}\natexlab{}.
\newblock \showarticletitle{{PUG:} a framework and practical implementation for
  why and why-not provenance}.
\newblock \bibinfo{journal}{\emph{{VLDB} J.}} (\bibinfo{year}{2019}),
  \bibinfo{pages}{47--71}.
\newblock


\bibitem[\protect\citeauthoryear{Lee, Lud{\"{a}}scher, and Glavic}{Lee
  et~al\mbox{.}}{2020}]%
        {DBLP:journals/pvldb/LeeLG20}
\bibfield{author}{\bibinfo{person}{Seokki Lee}, \bibinfo{person}{Bertram
  Lud{\"{a}}scher}, {and} \bibinfo{person}{Boris Glavic}.}
  \bibinfo{year}{2020}\natexlab{}.
\newblock \showarticletitle{Approximate Summaries for Why and Why-not
  Provenance}.
\newblock \bibinfo{journal}{\emph{Proc. {VLDB} Endow.}} (\bibinfo{year}{2020}),
  \bibinfo{pages}{912--924}.
\newblock


\bibitem[\protect\citeauthoryear{Li, Cornelius, Liu, Wang, and Barab{\'a}si}{Li
  et~al\mbox{.}}{2017}]%
        {li2017fundamental}
\bibfield{author}{\bibinfo{person}{Aming Li}, \bibinfo{person}{Sean~P
  Cornelius}, \bibinfo{person}{Y-Y Liu}, \bibinfo{person}{Long Wang}, {and}
  \bibinfo{person}{A-L Barab{\'a}si}.} \bibinfo{year}{2017}\natexlab{}.
\newblock \showarticletitle{The fundamental advantages of temporal networks}.
\newblock  (\bibinfo{year}{2017}), \bibinfo{pages}{1042--1046}.
\newblock


\bibitem[\protect\citeauthoryear{Masuda and Holme}{Masuda and Holme}{2013}]%
        {masuda2013predicting}
\bibfield{author}{\bibinfo{person}{Naoki Masuda} {and} \bibinfo{person}{Petter
  Holme}.} \bibinfo{year}{2013}\natexlab{}.
\newblock \showarticletitle{Predicting and controlling infectious disease
  epidemics using temporal networks}.
\newblock \bibinfo{journal}{\emph{F1000Prime Reports}} \bibinfo{volume}{5},
  \bibinfo{number}{6} (\bibinfo{year}{2013}).
\newblock


\bibitem[\protect\citeauthoryear{Morris}{Morris}{1985}]%
        {MorrisTR}
\bibfield{author}{\bibinfo{person}{R.~T. Morris}.}
  \bibinfo{year}{1985}\natexlab{}.
\newblock \bibinfo{booktitle}{\emph{A Weakness in the 4.2BSD Unix TCP/IP
  Software}}.
\newblock \bibinfo{type}{{T}echnical {R}eport} \#117.
  \bibinfo{institution}{Bell Labs Computer Science}.
\newblock


\bibitem[\protect\citeauthoryear{Namaki, Floratou, Psallidas, Krishnan,
  Agrawal, Wu, Zhu, and Weimer}{Namaki et~al\mbox{.}}{2020}]%
        {DBLP:conf/kdd/NamakiFPKAWZW20}
\bibfield{author}{\bibinfo{person}{Mohammad~Hossein Namaki},
  \bibinfo{person}{Avrilia Floratou}, \bibinfo{person}{Fotis Psallidas},
  \bibinfo{person}{Subru Krishnan}, \bibinfo{person}{Ashvin Agrawal},
  \bibinfo{person}{Yinghui Wu}, \bibinfo{person}{Yiwen Zhu}, {and}
  \bibinfo{person}{Markus Weimer}.} \bibinfo{year}{2020}\natexlab{}.
\newblock \showarticletitle{Vamsa: Automated Provenance Tracking in Data
  Science Scripts}. In \bibinfo{booktitle}{\emph{{KDD} '20: The 26th {ACM}
  {SIGKDD} Conference on Knowledge Discovery and Data Mining, Virtual Event,
  CA, USA, August 23-27, 2020}}. \bibinfo{pages}{1542--1551}.
\newblock


\bibitem[\protect\citeauthoryear{Parulian, McPhillips, and
  Lud{\"{a}}scher}{Parulian et~al\mbox{.}}{2021}]%
        {DBLP:conf/ipaw/ParulianML21}
\bibfield{author}{\bibinfo{person}{Nikolaus~Nova Parulian},
  \bibinfo{person}{Timothy~M. McPhillips}, {and} \bibinfo{person}{Bertram
  Lud{\"{a}}scher}.} \bibinfo{year}{2021}\natexlab{}.
\newblock \showarticletitle{A Model and System for Querying Provenance from
  Data Cleaning Workflows}. In \bibinfo{booktitle}{\emph{Provenance and
  Annotation of Data and Processes - 8th and 9th International Provenance and
  Annotation Workshop, {IPAW} 2020 + {IPAW} 2021, Virtual Event, July 19-22,
  2021, Proceedings}}. \bibinfo{pages}{183--197}.
\newblock


\bibitem[\protect\citeauthoryear{Polychroniou, Raghavan, and Ross}{Polychroniou
  et~al\mbox{.}}{2015}]%
        {polychroniou2015rethinking}
\bibfield{author}{\bibinfo{person}{Orestis Polychroniou}, \bibinfo{person}{Arun
  Raghavan}, {and} \bibinfo{person}{Kenneth~A Ross}.}
  \bibinfo{year}{2015}\natexlab{}.
\newblock \showarticletitle{Rethinking SIMD vectorization for in-memory
  databases}. In \bibinfo{booktitle}{\emph{Proceedings of the 2015 ACM SIGMOD
  International Conference on Management of Data}}.
  \bibinfo{pages}{1493--1508}.
\newblock


\bibitem[\protect\citeauthoryear{Psallidas and Wu}{Psallidas and Wu}{2018a}]%
        {DBLP:conf/sigmod/Psallidas018}
\bibfield{author}{\bibinfo{person}{Fotis Psallidas} {and}
  \bibinfo{person}{Eugene Wu}.} \bibinfo{year}{2018}\natexlab{a}.
\newblock \showarticletitle{Demonstration of Smoke: {A} Deep Breath of
  Data-Intensive Lineage Applications}. In
  \bibinfo{booktitle}{\emph{Proceedings of the 2018 International Conference on
  Management of Data, {SIGMOD} Conference 2018, Houston, TX, USA, June 10-15,
  2018}}. \bibinfo{pages}{1781--1784}.
\newblock


\bibitem[\protect\citeauthoryear{Psallidas and Wu}{Psallidas and Wu}{2018b}]%
        {DBLP:journals/pvldb/PsallidasW18}
\bibfield{author}{\bibinfo{person}{Fotis Psallidas} {and}
  \bibinfo{person}{Eugene Wu}.} \bibinfo{year}{2018}\natexlab{b}.
\newblock \showarticletitle{Smoke: Fine-grained Lineage at Interactive Speed}.
\newblock \bibinfo{journal}{\emph{Proc. {VLDB} Endow.}} (\bibinfo{year}{2018}),
  \bibinfo{pages}{719--732}.
\newblock


\bibitem[\protect\citeauthoryear{Richardson and Domingos}{Richardson and
  Domingos}{2002}]%
        {DBLP:conf/kdd/RichardsonD02}
\bibfield{author}{\bibinfo{person}{Matthew Richardson} {and}
  \bibinfo{person}{Pedro~M. Domingos}.} \bibinfo{year}{2002}\natexlab{}.
\newblock \showarticletitle{Mining knowledge-sharing sites for viral
  marketing}. In \bibinfo{booktitle}{\emph{Proceedings of the Eighth {ACM}
  {SIGKDD} International Conference on Knowledge Discovery and Data Mining}}.
  \bibinfo{pages}{61--70}.
\newblock


\bibitem[\protect\citeauthoryear{Ruan, Chen, Dinh, Lin, Ooi, and Zhang}{Ruan
  et~al\mbox{.}}{2019}]%
        {ruan2019fine}
\bibfield{author}{\bibinfo{person}{Pingcheng Ruan}, \bibinfo{person}{Gang
  Chen}, \bibinfo{person}{Tien Tuan~Anh Dinh}, \bibinfo{person}{Qian Lin},
  \bibinfo{person}{Beng~Chin Ooi}, {and} \bibinfo{person}{Meihui Zhang}.}
  \bibinfo{year}{2019}\natexlab{}.
\newblock \showarticletitle{Fine-grained, secure and efficient data provenance
  on blockchain systems}.
\newblock \bibinfo{journal}{\emph{Proceedings of the VLDB Endowment}}
  (\bibinfo{year}{2019}), \bibinfo{pages}{975--988}.
\newblock


\bibitem[\protect\citeauthoryear{Rupprecht, Davis, Arnold, Gur, and
  Bhagwat}{Rupprecht et~al\mbox{.}}{2020}]%
        {DBLP:journals/pvldb/RupprechtDAGB20}
\bibfield{author}{\bibinfo{person}{Lukas Rupprecht}, \bibinfo{person}{James~C.
  Davis}, \bibinfo{person}{Constantine Arnold}, \bibinfo{person}{Yaniv Gur},
  {and} \bibinfo{person}{Deepavali Bhagwat}.} \bibinfo{year}{2020}\natexlab{}.
\newblock \showarticletitle{Improving Reproducibility of Data Science Pipelines
  through Transparent Provenance Capture}.
\newblock \bibinfo{journal}{\emph{Proc. {VLDB} Endow.}} (\bibinfo{year}{2020}),
  \bibinfo{pages}{3354--3368}.
\newblock


\bibitem[\protect\citeauthoryear{Savage, Wetherall, Karlin, and
  Anderson}{Savage et~al\mbox{.}}{[n.d.]}]%
        {DBLP:conf/sigcomm/SavageWKA00}
\bibfield{author}{\bibinfo{person}{Stefan Savage}, \bibinfo{person}{David
  Wetherall}, \bibinfo{person}{Anna~R. Karlin}, {and}
  \bibinfo{person}{Thomas~E. Anderson}.} \bibinfo{year}{[n.d.]}\natexlab{}.
\newblock \showarticletitle{Practical network support for {IP} traceback}. In
  \bibinfo{booktitle}{\emph{Proceedings of the {ACM} {SIGCOMM} 2000 Conference
  on Applications, Technologies, Architectures, and Protocols for Computer
  Communication, August 28 - September 1, 2000, Stockholm, Sweden}}.
  \bibinfo{pages}{295--306}.
\newblock


\bibitem[\protect\citeauthoryear{Taxidou, Nies, Verborgh, Fischer, Mannens, and
  de~Walle}{Taxidou et~al\mbox{.}}{2015}]%
        {DBLP:conf/www/TaxidouNVFMW15}
\bibfield{author}{\bibinfo{person}{Io Taxidou}, \bibinfo{person}{Tom~De Nies},
  \bibinfo{person}{Ruben Verborgh}, \bibinfo{person}{Peter~M. Fischer},
  \bibinfo{person}{Erik Mannens}, {and} \bibinfo{person}{Rik~Van de Walle}.}
  \bibinfo{year}{2015}\natexlab{}.
\newblock \showarticletitle{Modeling Information Diffusion in Social Media as
  Provenance with {W3C} {PROV}}. In \bibinfo{booktitle}{\emph{Proceedings of
  the 24th International Conference on World Wide Web Companion, {WWW}}}.
  \bibinfo{pages}{819--824}.
\newblock


\bibitem[\protect\citeauthoryear{Zhou, Zheng, Han, and He}{Zhou
  et~al\mbox{.}}{2020}]%
        {DBLP:conf/kdd/ZhouZ0H20}
\bibfield{author}{\bibinfo{person}{Dawei Zhou}, \bibinfo{person}{Lecheng
  Zheng}, \bibinfo{person}{Jiawei Han}, {and} \bibinfo{person}{Jingrui He}.}
  \bibinfo{year}{2020}\natexlab{}.
\newblock \showarticletitle{A Data-Driven Graph Generative Model for Temporal
  Interaction Networks}. In \bibinfo{booktitle}{\emph{{KDD} '20: The 26th {ACM}
  {SIGKDD} Conference on Knowledge Discovery and Data Mining, Virtual Event,
  CA, USA, August 23-27, 2020}}. \bibinfo{pages}{401--411}.
\newblock


\bibitem[\protect\citeauthoryear{Z{\"{u}}fle, Renz, Emrich, and
  Franzke}{Z{\"{u}}fle et~al\mbox{.}}{2018}]%
        {DBLP:conf/edbt/ZufleREF18}
\bibfield{author}{\bibinfo{person}{Andreas Z{\"{u}}fle},
  \bibinfo{person}{Matthias Renz}, \bibinfo{person}{Tobias Emrich}, {and}
  \bibinfo{person}{Maximilian Franzke}.} \bibinfo{year}{2018}\natexlab{}.
\newblock \showarticletitle{Pattern Search in Temporal Social Networks}. In
  \bibinfo{booktitle}{\emph{Proceedings of the 21st International Conference on
  Extending Database Technology, {EDBT} 2018, Vienna, Austria, March 26-29,
  2018}}. \bibinfo{pages}{289--300}.
\newblock


\end{thebibliography}

\end{document}